\documentclass[%
superscriptaddress,
nofootinbib,
 amsmath,amssymb,
 aps, 
prd,
]{revtex4-2}
\usepackage{graphicx} % Required for inserting images
\usepackage{dcolumn}% Align table columns on decimal point
\usepackage{bm}% bold math
\usepackage{hyperref}% add hypertext capabilities
\usepackage{physics}
\usepackage{xcolor}
\usepackage{tikz}
\usetikzlibrary{arrows.meta}
\usetikzlibrary{decorations.pathreplacing}
\usepackage{placeins}
\usepackage{orcidlink}
\usepackage[T1]{fontenc}

\begin{document}

\title{Tightening energy-based boson truncation bound using Monte Carlo-assisted methods}

\author{Jinghong Yang$^{\orcidlink{0009-0003-8844-8482}}$}
\email{yangjh@umd.edu}
\affiliation{Department of Physics, University of Maryland, College Park, MD 20742, USA}
\affiliation{National Quantum Lab, University of Maryland, College Park, MD 20742, USA}
\author{Christopher F. Kane$^{\orcidlink{0000-0002-2020-8971}}$}
\email{cfkane24@umd.edu}
\affiliation{Department of Physics, University of Maryland, College Park, MD 20742, USA}
\affiliation{Maryland Center for Fundamental Physics, University of Maryland, College Park, MD 20742, USA}
\affiliation{Joint Center for Quantum Information and Computer Science, University of Maryland, College Park, Maryland 20742, USA.}
\author{Shabnam Jabeen$^{\orcidlink{0000-0002-0155-7383}}$}
\email{jabeen@umd.edu}
\affiliation{Department of Physics, University of Maryland, College Park, MD 20742, USA}
\affiliation{National Quantum Lab, University of Maryland, College Park, MD 20742, USA}
\begin{abstract}

Quantum simulation offers a promising framework for quantum field theory calculations. Obtaining reliable results, however, requires careful characterization of systematic uncertainties. One important source is the boson truncation error, which arises from representing infinite-dimensional local Hilbert spaces with finite-dimensional ones. Previous studies have examined this problem from several perspectives. In particular, Ref.~\cite{jordan_quantum_2012} derived an energy-based bound applicable to generic low-energy states across a broad class of field theories. However, this approach often yields overly conservative bounds, especially at large volumes. In this work, we introduce a new methodology that significantly tightens the energy-based boson truncation bound through two complementary advances: an improved analytic derivation and a Monte Carlo-based numerical procedure. We demonstrate the method in (1+1)-dimensional scalar field theory and (2+1)-dimensional U(1) gauge theory in the dual formalism. Our approach substantially mitigates the volume dependence of the required truncation cutoff, achieving reductions nearly proportional to the volume in some cases and to the square root of the volume in others.

\end{abstract}

\maketitle

\tableofcontents

\section{Introduction}

The study of quantum field theory (QFT) is central to many areas of physics, including particle physics~\cite{peskin_introduction_1995}, condensed matter~\cite{shankar_quantum_2017}, and cosmology~\cite{mukhanov_physical_2005}. Consequently, our understanding of many physical phenomena relies heavily on the ability to perform robust QFT calculations. While perturbative methods have been highly successful in certain regimes~\cite{ParticleDataGroup:2024cfk,aliberti_anomalous_2025}, many of the most interesting phenomena arise in strongly coupled settings where such approaches break down~\cite{wilson_confinement_1974,fodor_light_2012,guenther_overview_2021}.
In these cases, non-perturbative methods become essential. Among these, lattice field theory has emerged as one of the most important computational frameworks~\cite{wilson_confinement_1974,gattringer_quantum_2010}.
In this approach, spacetime is restricted to a finite volume and discretized into lattice points.
Field configurations in the Euclidean path integral are then represented by multidimensional arrays 
defined on the lattice, enabling numerical calculation by Monte Carlo sampling~\cite{gattringer_quantum_2010}.
As a systematically improvable approach with controllable uncertainties,  lattice field theory has been successfully applied to the calculation of hadron spectra~\cite{fodor_light_2012}, hadron structure~\cite{ji_largemomentum_2021,latticepartoncollaboration_nucleon_2023}, low-energy scattering amplitudes~\cite{briceno_scattering_2018}, and many other quantities~\cite{borsanyi_leading_2021,guenther_overview_2021,boccaletti_hybrid_2026,aoki_flag_2026}.

Despite the remarkable successes of traditional Euclidean lattice field theory, the methodology faces fundamental limitations.
In particular,  real-time quantum dynamics and finite-density calculations remain notoriously difficult, largely due to the sign problem in Monte Carlo importance sampling~\cite{troyer_computational_2005,goy_sign_2017,nagata_finitedensity_2022,wiese_quantum_2014,funcke_review_2023}.
Nonetheless, recent advances in quantum computing have opened a promising alternative avenue~\cite{byrnes_simulating_2006,jordan_quantum_2012,funcke_review_2023,bauer_quantum_2023,bauer_quantum_2023b}.
By discretizing space into lattice sites and mapping them onto a quantum register, the quantum state of the field can be approximately encoded into the Hilbert space of a quantum computer~\cite{abrams_simulation_1997,somma_quantum_2003,byrnes_simulating_2006,jordan_quantum_2012,jordan_quantum_2014}.
Given a Hamiltonian, the corresponding time evolution can often be efficiently simulated using quantum circuits, allowing access to real-time dynamical quantities~\cite{lloyd_universal_1996,nielsen_quantum_2000,somma_quantum_2003,byrnes_simulating_2006,jordan_quantum_2012}.
This approach is commonly referred to as the \textit{Hamiltonian formulation} or \textit{Minkowski lattice field theory}~\cite{kogut_hamiltonian_1975,gustafson_quantum_2021}, in contrast to the traditional \textit{Lagrangian formulation} or \textit{Euclidean lattice field theory} implemented on classical computers. 
A seminal early work in this direction is the study by Jordan, Lee, and Preskill \cite{jordan_quantum_2012}, which proposed a quantum algorithm for simulating scattering processes in scalar field theory.
Since then, quantum simulation has developed rapidly and has been applied to a wide range of problems, including particle scattering~\cite{jordan_quantum_2012,jordan_quantum_2014,jordan_quantum_2014a,pedernales_efficient_2014,lamm_general_2019,bender_gauge_2020,surace_scattering_2021,gustafson_realtime_2021,kreshchuk_simulating_2023,li_scattering_2024,belyansky_highenergy_2024,turco_quantum_2024,wu_efficient_2024,briceno_coherent_2024,sharma_scattering_2024,su_coldatom_2024,chai_fermionic_2025,ingoldby_realtime_2025,davoudi_quantum_2025,bennewitz_simulating_2025,farrell_digital_2025,zemlevskiy_scalable_2025,joshi_probing_2025,xiang_realtime_2025,guo_scattering_2026,guo_scattering_2026a,barata_hadronic_2025,chai_simulating_2026,hite_improved_2026}, 
hadron structure observables~\cite{mueller_deeply_2020,mueller_computing_2021,lamm_parton_2020,echevarria_quantum_2021,li_partonic_2022,li_exploring_2023,gustin_generalized_2023,grieninger_quasiparton_2024,li_simulating_2024,banuls_parton_2026,kang_partonic_2025,chen_parton_2025,ikeda_quantum_2025}, collective neutrino dynamics~\cite{hall_simulation_2021,jha_quantum_2022,amitrano_trappedion_2023,balantekin_quantum_2023,siwach_collective_2023,chernyshev_quantum_2024,singh_simulating_2024,turro_qutrit_2025,spagnoli_collective_2025,tripathi_quantum_2025,bleau_quantum_2026}, and far-from-equilibrium phenomena such as string breaking~\cite{mueller_quantum_2023,de_observation_2024,surace_stringbreaking_2024,davoudi_quantum_2024,luo_quantum_2025,alexandrou_realizing_2025,halimeh_quantum_2025,cataldi_realtime_2025,mueller_quantum_2025,halimeh_quantum_2025,grieninger_quantum_2026,cao_string_2026,gupta_stringbreaking_2026,joshi_observation_2026,xu_observation_2026}, among others~\cite{muschik_u1_2017,ciavarella_algorithm_2020,bauer_simulating_2021,choi_rodeo_2021,halimeh_gauge_2022,liu_variational_2022,barata_medium_2022,charles_simulating_2023,barata_quantum_2023a,lee_liouvillian_2023,khodaeva_quantum_2023,davoudi_scattering_2024,kane_nearlyoptimal_2024,bazavov_efficient_2024,barata_realtime_2025,li_framework_2025,gong_digit_2025,azad_barrenplateau_2025,lee_quantum_2025,qian_efficient_2025,araz_state_2025,chernyshev_pathfinding_2026,bombieri_u1_2026,ilcic_observation_2026,fulgado-claudio_hybrid_2026}.

Extracting accurate and reliable results from lattice field theory requires a careful analysis of systematic uncertainties.
In classical Euclidean lattice calculations, notable sources of systematic uncertainties include discretization effects due to finite lattice spacing~\cite{symanzik_continuum_1983,symanzik_continuum_1983a} and finite-volume effects arising from the restriction to a finite Euclidean spacetime box~\cite{luscher_volume_1986,luscher_volume_1986a}.
Both theoretical analysis and numerical extrapolation have been employed to control these uncertainties~\cite{symanzik_continuum_1983,symanzik_continuum_1983a,luscher_volume_1986,luscher_volume_1986a,durr_abinitio_2008,aoki_review_2017,constantinou_lattice_2022}.
Likewise, establishing quantum simulation as a reliable computational framework requires a similarly thorough characterization of its systematic uncertainties.
These include implementation-level errors, such as Trotterization errors~\cite{lloyd_universal_1996,childs_theory_2021}, shot noise~\cite{abrams_simulation_1997,knill_optimal_2007}, and hardware noise~\cite{chen_benchmarking_2024,abughanem_ibm_2025,ransford_helios_2025,bluvstein_faulttolerant_2026}, as well as physical systematic effects, e.g., the aforementioned finite lattice spacing and finite volume errors~\cite{luo_improved_1999,carlsson_direct_2001,carena_lattice_2021,carena_gauge_2022,carena_improved_2022,clemente_strategies_2022,crippa_determining_2024,kane_obtaining_2025,gross_matching_2025,schwagerl_fermion_2025,illa_dynamical_2025,danna_adiabatic_2025,illa_improved_2025,wang_bounding_2021,briceno_role_2021,burbano_realtime_2026}.
For quantum simulation, an additional physical systematic error arises due to the digitization of the boson fields.
Because the bosonic field degree of freedom at each lattice site resides in an infinite-dimensional Hilbert space, practical simulation requires a \textit{digitization} or \textit{truncation} into a finite-dimensional space~\cite{jordan_quantum_2012,byrnes_simulating_2006}.
The error induced by this truncation must therefore be quantified in order to establish controlled and reliable quantum simulation results.

Various methods have been proposed to bound and analyze this digitization error. The earliest quantitative discussion of this problem appears in Ref. \cite{jordan_quantum_2012}, which derives an energy-based bound showing that the truncation error is suppressed polynomially as the truncation cutoff increases.
Since then, a variety of alternative strategies have emerged, including bounds based on quantum speed limit \cite{tong_provably_2022}, perturbation theory \cite{ciavarella_truncation_2025}, Nyquist-Shannon sampling theorem~\cite{macridin_electronphonon_2018,macridin_digital_2018,klco_digitization_2019,macridin_bosonic_2022}, Hamiltonian truncation effective theory~\cite{schmoll_hamiltonian_2023,ingoldby_enhancing_2024,draper_hamiltonian_2025}, and others~\cite{hackett_digitizing_2019,alexandru_gluon_2019,hanada_estimating_2023,calliari_field_2026}. 
Each approach possesses its own strengths and regimes of applicability. 
Notably, while the energy-based bound in Ref.~\cite{jordan_quantum_2012} can often be overly conservative,
it remains highly relevant due to its broad applicability~\cite{liu_quantum_2021,buser_quantum_2021,Watson:2023oov,yao_quantum_2025}. 

As the energy based bound in Ref.~\cite{jordan_quantum_2012} can be loose, it is highly desirable to derive a tighter estimate.
Reducing the required truncation cutoff is not only directly relevant for lowering qubit resource costs, but is also important for improving bounds on time-evolution errors.
For commonly used time evolution algorithms, such as Trotterization~\cite{childs_theory_2021} and quantum signal processing (QSP) \cite{Low:2016sck}, 
the resource requirements depend
on the norm of the Hamiltonian terms (or their commutators), which in turn can scale polynomially with the boson truncation cutoff.
Consequently, a tighter truncation bound can indirectly lead to 
reduced costs
for time evolution and related algorithms, such as adiabatic state preparation \cite{jordan_quantum_2012,danna_adiabatic_2025} and eigenstate filtering using QSP~\cite{lin_optimal_2020}.

In this work, we develop a method to tighten the energy-based bound introduced in Ref.~\cite{jordan_quantum_2012}.
Specifically, we introduce two novel techniques, one numerical and one analytical, which we refer to as the ``Monte Carlo trick'' and the ``$p$-norm trick,'' respectively.  
By combining these two complementary strategies, we obtain an improved boson truncation bound.
Our method substantially alleviates the volume dependence of the truncation cutoff, leading to significant reductions at large volumes.
We first present the method in the context of scalar field theory, following the setup originally considered in Ref.~\cite{jordan_quantum_2012}. 
To illustrate how this framework can be generalized to other theories, we further analyze the truncation bound for (2+1)-dimensional U(1) gauge theory in the dual formalism \cite{bender_gauge_2020,bauer_efficient_2021,kane_efficient_2022} as a concrete example.

This paper is organized as follows. In Sec.~\ref{sec:preview}, we provide a brief preview of the main results. In Sec.~\ref{sec:background}, we compare different approaches to bounding boson truncation error in the literature and review the energy-based bound from Ref.~\cite{jordan_quantum_2012}. Here, we also present an analysis of the origin of this bound's looseness and, in particular, identify an implicit square-root volume factor that is often overlooked in the literature. The following two sections describe our methodology.
Because the field variable (e.g., $\phi$) and its conjugate momentum field (e.g., $\pi$) require different treatments in our framework, we partition the discussion accordingly.
Sec.~\ref{sec:method} focuses on the treatment of the field variable, while Sec.~\ref{sec:methodII} addresses the conjugate momentum field. We then present the numerical results in Sec. \ref{sec:results}. Finally,  Sec.~\ref{sec:time_evolution} discusses the implications of our results for error analysis and resource estimates of time evolution, and Sec.~\ref{sec:conclude} concludes the paper.

\section{Summary of results\label{sec:preview}}

As a preview of the main results in this paper, we briefly summarize our key findings and present representative numerical examples. A discussion of the background and our methodology is provided in the later sections.

In their seminal work, Jordan, Lee, and Preskill \cite{jordan_quantum_2012} outlined a framework for using quantum simulation to calculate scattering amplitudes. To estimate the required truncation cutoff for field digitization, the authors employed an energy-based bound. While physically motivated and robust, this bound can be rather loose. In particular, it exhibits unfavorable scaling with respect to the lattice volume $\mathcal{V}$. As derived in Ref.~\cite{jordan_quantum_2012}, the truncation cutoffs for $\phi$ and $\pi$ fields to achieve an error $\epsilon$ scale as $\sqrt{\frac{\mathcal{V} E}{m_0^2 \epsilon}}$ and $\sqrt{\frac{\mathcal{V} E}{\epsilon}}$, respectively, both of which contain an explicit $\sqrt{\mathcal{V}}$ factor. Moreover, although this point is often overlooked, there is typically an additional implicit $\sqrt{\mathcal{V}}$ factor arising from $\sqrt{E}$, on which we will elaborate in later sections. Consequently, the original energy-based bound in Ref.~\cite{jordan_quantum_2012} scales poorly with the system volume.

In this work, we combine Monte Carlo calculations with analytic derivations to tighten the energy based bound.
Our improved method yields a cutoff $\phi_{\text{max}}$ with only a mild dependence on volume, appearing to scale logarithmically or algebraically with a small exponent, thus achieving nearly a factor of $\mathcal{V}$ improvement. For the $\pi$ field, the improved method can achieve a factor of $\sqrt{\mathcal{V}}$ improvement. As an illustration, we present representative results in Table~\ref{tab:phi4_demo}. Our techniques can also be extended to other theories.
To demonstrate its feasibility, we further discuss non-compact 2+1D U(1) gauge theory in the dual formalism, where similar improvements can be observed. Detailed results are presented in the later sections.

\begin{table}[]
    \centering
    \begin{tabular}{c|c|c|c|c}
         & $\phi_{\rm max}$ with $N_S=8$& $\pi_{\rm max}$ with $N_S=8$& $\phi_{\rm max}$ with $N_S=32$&$\pi_{\rm max}$ with $N_S=32$\\
         \hline
         Energy-based bounds~\cite{jordan_quantum_2012} & 236& 118& 882&441\\
         This work & 33& 44& 34&84\\
 Ratio& 7.2& 2.7& 26&5.2\\\end{tabular}
    \caption{Comparison of truncation cutoffs for field values $\phi_{\rm max}$ and conjugate field values $\pi_{\rm max}$ in a 1+1D scalar field theory required to achieve a digitization error $\epsilon = 0.01$ calculated using the energy-based methods in Ref.~\cite{jordan_quantum_2012} and the methods in this work.  $N_S$ is the number of lattice sites. The energy of the state $\ket{\psi}$ relative to the vacuum is assumed to be below $1/a$, where $a$  is the lattice spacing. 
    The bare parameters are $m_0=1/2$ and $\lambda_0=16$. Here, the physical quantities have been rescaled by powers of $a$ to make them dimensionless. See Sec.~\ref{sec:results:phi} for more details.
    }
    \label{tab:phi4_demo}
\end{table}

\section{Background\label{sec:background}}

\subsection{Brief review of literature\label{sec:background:literature}}
To study boson truncation errors, several studies have been conducted in addition to the foundational energy-based bound proposed in Ref.~\cite{jordan_quantum_2012}. These works approach the truncation problem from multiple perspectives; in this section, we provide a brief review of these developments and discuss why the energy-based bound remains a highly relevant method.

To begin with, the energy-based bound was introduced by Jordan, Lee, and Preskill in Ref.~\cite{jordan_quantum_2012}. 
For a quantum state under a given energy, their method produces a truncation error bound that decays polynomially with respect to the truncation cutoff.
Henceforth, we refer to this approach as either the energy-based bound or the JLP bound. This method is reviewed in further detail in Sec. \ref{sec:background:jlp}.

In Refs.~\cite{macridin_electronphonon_2018,macridin_digital_2018,macridin_bosonic_2022}, the authors employ the Nyquist-Shannon (NS) sampling theorem~\cite{shannon_communication_1949} to analyze boson truncation errors. Originally developed in signal processing, the NS theorem provides a framework for quantifying how accurately a continuous wave-function can be represented by a finite set of discrete samples. Within this approach, the authors prove that the truncation errors associated with the field variable and its conjugate momentum are exponentially suppressed as the cutoff is increased, provided the state has bounded boson occupation number (or sufficiently suppressed occupation-number tails). The formalism has been applied to state fidelity, low-lying eigenenergies, and corrections to the canonical commutation relations. However, the resulting error estimate is conditional on controlling the occupation-number distribution, which may itself be nontrivial. While this pre-condition can often be satisfied in electron-phonon systems, establishing it may be more challenging in more complicated cases \cite{bolz_higher_1998,brodsky_condensates_2011}, especially for field theory in  highly non-perturbative regions \cite{chabysheva_convergence_2019}.
A related work \cite{klco_digitization_2019} also draws inspiration from the NS theorem and includes a numerical analysis of the correction to the ground-state energy for small systems.

Another approach involves Hamiltonian truncation effective theory \cite{schmoll_hamiltonian_2023,ingoldby_enhancing_2024,draper_hamiltonian_2025}. In this framework, the Hamiltonian is typically expressed in the momentum basis and then truncated. The truncation error is estimated numerically for small systems, e.g., by exact diagonalization.

A different perspective is taken in Ref.~\cite{calliari_field_2026}, which studies the digitization of U(1) gauge theory as a $\mathbb{Z}_N$ model from a renormalization-group viewpoint, treating $N$ as a coupling parameter. Numerical results are obtained using tensor-network methods.

References~\cite{hackett_digitizing_2019,hanada_estimating_2023} employ Monte Carlo simulation to estimate the corrections to the vacuum expectation values of certain operators. This provides a scalable method for assessing the resource requirements of field truncation. It should be noted that while both these studies and our work use Euclidean path-integral Monte Carlo calculations, they serve different purposes. Refs.~\cite{hackett_digitizing_2019,hanada_estimating_2023} use Monte Carlo to estimate and compare ground-state properties before and after the truncation, whereas we employ it as an auxiliary tool to tighten the energy-based bound in Ref.~\cite{jordan_quantum_2012}, which is applicable to a generic quantum state under a given energy threshold.

Ref.~\cite{alexandru_gluon_2019} also uses classical Monte Carlo calculation, albeit for a different purpose. Rather than comparing the digitized theory and full theory at a fixed lattice spacing, the authors investigate whether a particular digitization strategy for SU(3) gauge theory can reproduce the continuum-extrapolated values of observables of the full theory.

Notably, Ref.~\cite{tong_provably_2022} proposes a rigorous method based on the Lieb-Robinson bound.
Assuming the initial state has a bounded quantum number, their method bounds the growth of that quantum number over time, with the error decaying exponentially as a function of the cutoff. 
In addition, a truncation bound is derived for Hamiltonian eigenstates in the presence of a mass gap. 
Ref.~\cite{ciavarella_truncation_2025} derives similar results using perturbation theory instead, achieving a factorial suppression of the error with respect to the truncation cutoff. 
While these two methods offer asymptotically better bounds than the energy-based approach in Ref.~\cite{jordan_quantum_2012}, 
they are restricted to specific classes of Hamiltonians. As explicitly stated in Ref.~\cite{tong_provably_2022}, their method applies to U(1) and SU(2) Hamiltonians in the electric basis but does not extend to $\phi^4$ theory. 
Some other theories also appear to fall outside their scope, such as the non-compact version of the dual formalism of U(1) gauge theory~\cite{bauer_efficient_2021}, the orbifold formulation of lattice field theory~\cite{buser_quantum_2021}, and lattice nuclear effective field theories~\cite{Watson:2023oov}.

In summary, the various methods in the literature each have distinct strengths and limitations. 
Although the JLP energy-based bound guarantees only polynomial convergence, it is applicable to a broad class of theories and can bound the state fidelity for a generic quantum state under a specified energy threshold. Due to its generality, this method remains a valuable and highly relevant approach for truncation error analysis. Thus, being able to tighten the energy-based bound would be beneficial to the community.

\subsection{Review of energy-based bound in \cite{jordan_quantum_2012} \label{sec:background:jlp}}
In this section, we review the derivation and results of the JLP energy-based bound~\cite{jordan_quantum_2012}. 
We then analyze the underlying reasons for the bound being over-conservative and qualitatively explain why it can be tightened.

In Ref.~\cite{jordan_quantum_2012}, the authors derived a bound for the $\phi^4$ scalar theory. The Hamiltonian has the form\footnote{More precisely, throughout this work, when lattice renormalization is taken into account, the factor $\frac{1}{a}$ appearing in Hamiltonians should be replaced by $\frac{c}{a}$, where $c$ is a parameter to be determined in scale setting~\cite{carena_lattice_2021,kane_obtaining_2025}.  This distinction does not affect the methods presented in this work.}
\begin{equation}
    \hat H = \frac{1}{a}\sum_{\mathbf{x}}  \left(\frac{1}{2} \hat \pi(\mathbf{x})^2+\frac{1}{2}(\mathbf{\nabla} \hat\phi)^2(\mathbf{x})+\frac{1}{2} m_0^2 \hat\phi(\mathbf{x})^2+\frac{\lambda_0}{4!}\hat\phi(\mathbf{x})^4\right).\label{eqn:phi4-H}
\end{equation}
where $d$ is the spatial dimension, $a$ is the lattice spacing, and $\mathbf \nabla$ denotes a discretized lattice derivative. Here, for convenience, we have rescaled the couplings $m_0$ and $\lambda_0$, as well as the field variables $\phi$ and $\pi$, by powers of $a$ to render them dimensionless. 
In this convention, we have the commutation relation $[\hat\phi(\mathbf{x}_i),\hat\pi(\mathbf{x}_j)]=i\delta_{ij}$.
For notational simplicity, we will use $\phi_i$ and $\phi(\mathbf{x}_i)$ interchangeably.

To analyze the truncation cutoff for $\phi$ for a given state $\ket{\psi}$ below a certain energy, they note that any state $\ket{\psi}$ can be expanded in the $\phi$ basis:
\begin{equation}
    \ket{\psi} = \int_{-\infty}^\infty d \phi_1 ...\int_{-\infty}^\infty d \phi_{\mathcal{V}} \psi(\phi_1, ..., \phi_\mathcal{V}) \ket{\phi_1, ..., \phi_\mathcal{V}},\label{eqn:state_in_phi_basis}
\end{equation}
where $\mathcal{V}$ is the total number of sites in the lattice.
In this case, $\rho(\phi_1, ..., \phi_\mathcal{V})=\left|\psi(\phi_1, ..., \phi_\mathcal{V})\right|^2$ describes the probability distribution of the $\phi$ field. To analyze the truncation error, one can let
\begin{equation}
    \ket{\psi_{\phi_\text{max}}} = \int_{-\phi_\text{max}}^{\phi_\text{max}} d \phi_1 ...\int_{-\phi_\text{max}}^{\phi_\text{max}} d \phi_{\mathcal{V}} \psi(\phi_1, ..., \phi_\mathcal{V}) \ket{\phi_1, ..., \phi_\mathcal{V}},
\end{equation}
which means
\begin{equation}
    \bra{\psi}\ket{\psi_{\phi_\text{max}}} = \int_{-\phi_\text{max}}^{\phi_\text{max}} d \phi_1 ...\int_{-\phi_\text{max}}^{\phi_\text{max}} d \phi_{\mathcal{V}}\rho(\phi_1, ..., \phi_\mathcal{V}).\label{eqn:trunc_state_overlap}
\end{equation}
The truncation error can be quantified as $\epsilon = 1-  \bra{\psi}\ket{\psi_{\phi_\text{max}}} $. As noted in Ref.~\cite{jordan_quantum_2012}, from Eqn. \ref{eqn:trunc_state_overlap}, the truncation error $\epsilon$ is the probability that at least one of $|\phi_1|$, $|\phi_2|$, ..., $|\phi_\mathcal{V}|$ exceeds $\phi_\text{max}$.

To upper bound this probability for 
 a single site $\mathbf{x}$, Ref.~\cite{jordan_quantum_2012} uses Chebyshev's inequality. Suppose one can evaluate the mean and standard deviation of $\phi$:
\begin{equation}
    \begin{aligned}
        \mu_{\phi(\mathbf{x})} =& \bra{\psi}\hat \phi(\mathbf{x})\ket{\psi} \\
        \sigma_{\phi(\mathbf{x})} =& \sqrt{\bra{\psi}\hat\phi^2(\mathbf{x})\ket{\psi}-\bra{\psi}\hat\phi(\mathbf{x})\ket{\psi}^2},
    \end{aligned}
\end{equation}
then, the probability for $|\phi(\mathbf{x})|$ to exceed the cutoff $ \phi_\text{max} = |\mu_{\phi(\mathbf{x})}|+c\sigma_{\phi(\mathbf{x})}$ is upper bounded by $1/c^2$. Notice, both $|\mu_{\phi(\mathbf{x})}|$ and $\sigma_{\phi(\mathbf{x})}$ are upper bounded by the expectation value $\sqrt{\bra{\psi}\hat\phi^2(\mathbf{x})\ket{\psi}}$. Therefore, the probability for single-site field value $|\phi(\mathbf{x})|$ to exceed the cutoff $\phi_\text{max}$ 
is bounded by
\begin{equation}
    \frac{\bra{\psi}\hat\phi^2(\mathbf{x})\ket{\psi}}{\left(\phi_\text{max}-\sqrt{\bra{\psi}\hat\phi^2(\mathbf{x})\ket{\psi}}\right)^2}.
\end{equation}
Since there are $\mathcal{V}$ sites in total, if $\bra{\psi}\hat\phi^2(\mathbf{x})\ket{\psi}$ can be bounded at any site, one can bound the total error $\epsilon$ as 
\begin{equation}
    \epsilon \leq \frac{\mathcal{V}\bra{\psi}\hat\phi^2(\mathbf{x})\ket{\psi}}{\left(\phi_\text{max}-\sqrt{\bra{\psi}\hat\phi^2(\mathbf{x})\ket{\psi}}\right)^2},
  %  \sim  \frac{\mathcal{V}\bra{\psi}\hat\phi^2(\mathbf{x})\ket{\psi}}{\left(\phi_\text{max}\right)^2}
  \label{eqn:jlp_err_tot}
\end{equation}
where the union bound $\Pr(A\cup B)\leq \Pr(A)+\Pr(B)$ has been used. In other words, given an error budget $\epsilon$, the cutoff $\phi_\text{max}$ needs to scale as 
\begin{equation}
    \phi_\text{max} = \left(\sqrt{\frac{\mathcal{V}}{\epsilon}}+1\right) \sqrt{\bra{\psi}\hat\phi^2(\mathbf{x})\ket{\psi}}
    \sim \sqrt{\frac{\mathcal{V}\bra{\psi}\hat\phi^2(\mathbf{x})\ket{\psi}}{\epsilon}}.\label{eqn:jlp-phi-scaling}
\end{equation}

Now, the question is reduced to bounding the expectation value $\bra{\psi}\hat \phi^2(\mathbf{x})\ket{\psi}$, which is achieved by Ref.~\cite{jordan_quantum_2012} using an energy-based approach. The authors assume that the energy of state is below a given value $\bra{\psi} \hat H \ket{\psi} \leq E$.  Assuming $m_0^2>0$ and $\lambda_0\geq0$ in Eqn. \ref{eqn:phi4-H},  every term in the Hamiltonian would be positive-semidefinite. In this case, one can write
\begin{equation}
    \begin{aligned}
        E \geq & \bra{\psi} \hat H \ket{\psi} \\
                =   & \bra{\psi}\frac{1}{a} \sum_{\mathbf{x}} \left(\frac{1}{2} \hat\pi(\mathbf{x})^2+\frac{1}{2}(\mathbf{\nabla} \hat\phi)^2(\mathbf{x})+\frac{1}{2} m_0^2 \hat\phi(\mathbf{x})^2+\frac{\lambda_0}{4!}\hat\phi(\mathbf{x})^4\right) \ket{\psi} \\ 
             \geq & \frac{1}{a}\bra{\psi} \frac{1}{2} m_0^2 \hat \phi(\mathbf{x}_i)^2\ket{\psi},
    \end{aligned}
    \label{eqn:jlp-energy}
\end{equation}
for any site $\mathbf{x}_i$. The last equality is obtained by dropping all terms except the $ \frac{1}{2} m_0^2 \hat\phi(\mathbf{x}_i)^2$ term. In this way, one can obtain an upper bound for $\bra{\psi}\hat\phi^2(\mathbf{x})\ket{\psi}$. The cutoff $\phi_\text{max}$ can thus be chosen accordingly. Analogously, one can bound $\bra{\psi}\hat\pi^2(\mathbf{x})\ket{\psi}$ and use it to choose a cutoff $\pi_\text{max}$ for the $\pi$ field, which is equivalent to choosing a discretization unit $\delta_\phi$ for the $\phi$ field. While Ref.~\cite{jordan_quantum_2012} only discusses the scalar field theory, the same idea can be applied to more general cases,\footnote{Even if the Hamiltonian terms may not be manifestly positive-semidefinite, one can typically transform the Hamiltonian into such a form by shifting the energy and applying certain manipulations \cite{yao_quantum_2025}. 
% \CFK{is it easy to say in one or two words the manipulations you are referring to?}
} and has been used or mentioned in Refs.~\cite{tong_provably_2022,liu_quantum_2021,buser_quantum_2021,Watson:2023oov,yao_quantum_2025}.

While the energy-based approach manages to obtain a bound for $\bra{\psi}\hat\phi^2(\mathbf{x})\ket{\psi}$, this bound is not tight for two reasons.
The first reason is that many terms in Eqn.~\ref{eqn:jlp-energy} are simply ``discarded.''
Namely, Ref.~\cite{jordan_quantum_2012} essentially writes the Hamiltonian $\hat H$ as two parts: $\hat H=\frac{1}{a} \frac{1}{2} m_0^2 \hat\phi(\mathbf{x}_i)^2+\hat H_\text{others}$. Then, the authors use the positive-semidefinite property of $\hat H_\text{others}$ to  obtain $\bra{\psi} \hat H_\text{others} \ket{\psi} \geq 0$, thus concluding that
\begin{equation}
    \bra{\psi} \frac{1}{2} m_0^2 \hat\phi(\mathbf{x}_i)^2\ket{\psi} = \bra{\psi} a \hat H \ket{\psi} - \bra{\psi} a \hat H_\text{others} \ket{\psi} \leq a E- 0 =a E. \label{eqn:throw-away-H_other}
\end{equation}
The above inequality is loose in the sense that $\bra{\psi} \hat H_\text{others} \ket{\psi}$ can in fact be a very large number, and using $0$ as its lower bound is very ``wasteful.''

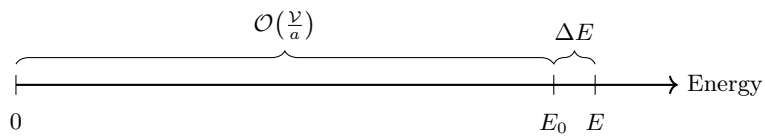
\begin{figure}[h]
    \centering
    \begin{tikzpicture}[scale=1.1]
    \draw[->, thick] (0,0) -- (8,0) node[right] {Energy};

    \def\xzero{0}
    \def\xEzero{6.5}
    \def\xE{7.0}

    \draw (\xzero,0.12) -- (\xzero,-0.12) node[below=4pt] {$0$};
    \draw (\xEzero,0.12) -- (\xEzero,-0.12) node[below=4pt] {$E_0$};
    \draw (\xE,0.12) -- (\xE,-0.12) node[below=4pt] {$E$};

    \draw[decorate,decoration={brace,amplitude=5pt}]
        (\xzero, 0.25) -- (\xEzero,0.25)
        node[midway,above=6pt]
        {$\mathcal{O}\!\left(\frac{\mathcal{V}}{a}\right)$};

    \draw[decorate,decoration={brace,amplitude=5pt}]
        (\xEzero,0.25) -- (\xE,0.25)
        node[midway,above=6pt] {$\Delta E$};
\end{tikzpicture}
    \caption{A schematic representation of the energy scales in the system. The vacuum energy $E_0$ scales as $\mathcal{O}(\mathcal{V}/a)$, whereas $\Delta E$ is significantly smaller in the typical scattering scenarios.}
    \label{fig:vacuum_energy_visual}
\end{figure}

To obtain a better qualitative understanding, note that the energy $E=\bra{\psi} \hat H \ket{\psi}$ can be written as two pieces: $E = E_0+\Delta E$, where $E_0$ is the ground state energy, also known as the vacuum energy, and $\Delta E$ is the energy relative to the ground state. 
When people discuss the energy scale of a scattering process, it is usually $\Delta E$ that is referred to. Compared to $E_0$, $\Delta E$ is typically much smaller (see Fig.~\ref{fig:vacuum_energy_visual} for a visualization). 
This is because, in lattice simulations of scattering, the particle momenta are limited by the finite lattice spacing, which acts as an ultraviolet regulator at the scale of $1/a$; thus, $\Delta E $ should remain moderate or small compared to this scale.
In contrast, the vacuum energy $E_0$ can be quite large. 
In the continuum or the infinite volume limit, $E_0$ approaches infinity. On a lattice, $E_0$ scales as $\mathcal{V} / a$, where $\mathcal{V}$ is the total number of sites. Thus, at large $\mathcal{V}$, $E\approx E_0$ is of order $\mathcal{O}(\mathcal{V}/a)$, making it much larger than $\Delta E$. Likewise, as $\hat H_\text{others}$ consists of $\mathcal{O}(\mathcal{V})$ terms, $\bra{\psi} \hat H_\text{others} \ket{\psi}$ is expected to scale as $\mathcal{O}(\mathcal{V}/a)$. On the other hand, as will be shown later in this paper, the difference between $ \bra{\psi} \hat H \ket{\psi}$ and $\bra{\psi} \hat H_\text{others} \ket{\psi}$ could be much smaller than $\mathcal{O}(\mathcal{V}/a)$. Thus, dropping the $\bra{\psi} \hat H_\text{others} \ket{\psi}$ term in Eqn. \ref{eqn:throw-away-H_other} will result in a bound that is much looser than needed. As we will show in this paper, it is possible to obtain a tighter lower bound for $\bra{\psi} \hat H_\text{others} \ket{\psi}$ using numerical methods.

The second source of looseness in JLP's energy-based bound is the lack of information regarding the distribution $|\psi(\phi_1,\dots, \phi_\mathcal{V})|^2$ of the field $\phi$, which is in general difficult to determine analytically. To overcome this limitation, Ref.~\cite{jordan_quantum_2012} employed Chebyshev's inequality, which is valid for any distribution $|\psi(\phi_1,\dots, \phi_\mathcal{V})|^2$, leading to $\epsilon \sim \phi_{\rm max}^{-2}$ in Eqn.~\ref{eqn:jlp_err_tot}.
Since the truncation error $\epsilon$ can scale linearly as the volume $\mathcal{V}$, the cutoff $\phi_\text{max}$ needs to increase accordingly to compensate, thereby introducing an explicit $\sqrt{\mathcal{V}}$ factor. 
Because Chebyshev's inequality yields a conservative bound, this $\sqrt{\mathcal{V}}$ factor overestimates the volume dependence of $\phi_\text{max}$.
To understand how knowledge of $|\psi(\phi_1,\dots, \phi_\mathcal{V})|^2$ could reduce the bound for $\phi_{\rm max}$, consider the simplified example where $|\psi(\phi_1,\dots, \phi_\mathcal{V})|^2 \sim \exp(-\sum_j \phi^2_j)$.
In this case, the error bound in Eqn.~\ref{eqn:jlp_err_tot} would become $\epsilon \sim e^{-\phi_{\rm max}^2}$, implying $\phi_{\rm max} \sim \text{polylog}(\mathcal{V})$, which represents a milder volume dependence.
While detailed analytic information about the distribution $|\psi(\phi_1,\dots, \phi_\mathcal{V})|^2$ is generally unavailable in practice, such information is not strictly necessary to tame the volume dependence of  $\phi_\text{max}$.
An improved scaling can be achieved by considering the global information about the probability distribution $|\psi(\phi_1,\dots, \phi_\mathcal{V})|^2$ rather than the distribution at each local site.
More specifically, we will later introduce a technique called the ``$p$-norm trick,'' which will enable us to incorporate more holistic information and derive better volume scaling, especially when combined with Monte Carlo methods.

\section{Methodology I: bounding the field variable \label{sec:method}}
Having argued in the previous section that the JLP energy-based bound admits considerable room for improvement, we now present our method to tighten the truncation bound.
To this end, we develop two techniques:
\begin{itemize}
    \item \textit{Monte Carlo trick}: a numerical Monte Carlo procedure for obtaining tighter estimates of key quantities;
    \item \textit{$p$-norm trick}: an analytical technique that exploits global information of the probability distribution, rather than treating each site independently.
\end{itemize}
Our method is best explained through examples. 
We begin with the scalar field theory and use it to illustrate how the bound for the $\phi$ field can be tightened in Sec.~\ref{sec:method:scalar}.
Specifically, we first introduce the ``Monte Carlo trick'' in Sec.~\ref{sec:method:scalar:basic}; then, in Sec.~\ref{sec:method:scalar:improved}, we proceed to explain the ``$p$-norm'' trick and how it can be combined with the ``Monte Carlo trick'' to yield a much tighter bound.
Then, to demonstrate that our approach can extend to more general settings, we derive a bound for the magnetic field in the dual formalism of the 2+1D non-compact U(1) gauge theory in Sec. \ref{sec:method:dual}.

\subsection{Scalar field theory\label{sec:method:scalar}}

\subsubsection{Basic idea: introducing ``Monte Carlo trick''\label{sec:method:scalar:basic}}
In this section, we will introduce the ``Monte Carlo trick.'' This trick allows us to derive a tighter upper bound for the expectation value $\bra{\psi}\hat\phi^2(\mathbf{x})\ket{\psi}$; plugging this back into Eqn.~\ref{eqn:jlp-phi-scaling}, one can obtain a tightened estimate for the truncation cutoff $\phi_\text{max}$.

Given a state $\ket{\psi}$ such that $\bra{\psi} \hat H \ket{\psi}\leq E$, we can write the energy $E$ in two parts: $E=E_0+\Delta E$, where $E_0$ is the vacuum energy and $\Delta E$ is the energy relative to the vacuum. 
In situations where people discuss the energy of a particle, it is the $\Delta E$ part that is commonly referred to.
That is to say, $\Delta E$ is the ``physical energy.'' For the energy-based bound, $\Delta E$ is the quantity that one can choose based on the kinematic setup of the scattering, while $E_0$ in general needs to be calculated numerically.

One can define $ \hat H'_{\phi(\mathbf{x}_i)}=\hat H - \frac{1}{2} m_0^2 \hat\phi(\mathbf{x}_i)^2$. Then, one has
\begin{equation}
    \begin{aligned}
        \bra{\psi} \frac{1}{2} m_0^2 \hat \phi(\mathbf{x}_i)^2\ket{\psi} =& \bra{\psi} a \hat H \ket{\psi} - \bra{\psi} a \hat H'_{\phi(\mathbf{x}_i)}\ket{\psi}.
    \end{aligned}\label{eqn:bound-improve-single-1}
\end{equation}
In Ref.~\cite{jordan_quantum_2012}, the authors use the positive-semidefinite property of $\hat H'_{\phi(\mathbf{x}_i)}$ to conclude that $\bra{\psi} \hat H'_{\phi(\mathbf{x}_i)}\ket{\psi}\geq 0$. Here, we will take an alternative approach. Notice, $\hat H'_{\phi(\mathbf{x}_i)}$ is a Hermitian operator. Let's denote its minimum eigenvalue as $E'_{0,\phi(\mathbf{x_i})}$. Then, we always have $\bra{\psi}\hat H'_{\phi(\mathbf{x}_i)}\ket{\psi}\geq E'_{0,\phi(\mathbf{x_i})}$. If we view the operator $\hat H'_{\phi(\mathbf{x}_i)}$ as a ``modified Hamiltonian,'' then $E'_{0,\phi(\mathbf{x_i})}$ can be regarded as the ground state energy of the modified Hamiltonian. As a ground state property, the energy $E'_{0,\phi(\mathbf{x_i})}$ is typically accessible through classical Monte Carlo simulation. Then, we can have 
\begin{equation}
    \bra{\psi} \frac{1}{2} m_0^2 \hat \phi(\mathbf{x}_i)^2\ket{\psi} \leq a(E_0-E'_{0,\phi(\mathbf{x_i})}) + a \Delta E.\label{eqn:scalar_bound_1}
\end{equation}
Here, compared to $E_0$, which scales with system volume $\mathcal{V}$,  the energy difference $ (E_0-E'_{0,\phi(\mathbf{x_i})})$ is expected to be much smaller. In this way, we can tighten the bound.

To evaluate or bound the energy difference $(E_0-E'_{0,\phi(\mathbf{x_i})})$, there are two methods. Firstly, one can evaluate $E_0$ and $E'_{0,\phi(\mathbf{x_i})}$ separately using Monte Carlo methods, as described in Appendix~\ref{app:path_integral:scalar}. 
Alternatively, one can bound $(E_0-E'_{0,\phi(\mathbf{x_i})})$  as follows.
For now, assume $\hat H'_{\phi(\mathbf{x}_i)}$ has a non-degenerate ground state and denote it as $\ket{\Omega_{\phi(\mathbf{x}_i)}}$; generalization to degenerate vacuum  should be possible.\footnote{In the derivation here, one can replace $\ket{\Omega_{\phi(\mathbf{x}_i)}}\bra{\Omega_{\phi(\mathbf{x}_i)}}$ with any mixed state $\hat\rho_{\Omega}$ supported on the ground-state subspace of $\hat H'_{\phi(\mathbf{x}_i)}$.}
By the definition of $E_0$, we must have 
\begin{equation}
\begin{aligned}
    E_0  = & \min_{\ket{\alpha}} \bra{\alpha} \hat{H} \ket{\alpha} \\
    \leq & \bra{\Omega_{\phi(\mathbf{x}_i)}} \hat H \ket{\Omega_{\phi(\mathbf{x}_i)}} \\
    =& \bra{\Omega_{\phi(\mathbf{x}_i)}}  \hat H'_{\phi(\mathbf{x}_i)}\ket{\Omega_{\phi(\mathbf{x}_i)}} + \bra{\Omega_{\phi(\mathbf{x}_i)}} \frac{1}{a}\frac{1}{2} m_0^2 \hat\phi(\mathbf{x}_i)^2\ket{\Omega_{\phi(\mathbf{x}_i)}}\\
    =& E'_{0,\phi(\mathbf{x_i})} + \bra{\Omega_{\phi(\mathbf{x}_i)}} \frac{1}{a}\frac{1}{2} m_0^2 \hat\phi(\mathbf{x}_i)^2\ket{\Omega_{\phi(\mathbf{x}_i)}}
\end{aligned}
\end{equation}
Therefore, we know
\begin{equation}
 a   \left(E_0 - E'_{0,\phi(\mathbf{x_i})}\right) \leq\bra{\Omega_{\phi(\mathbf{x}_i)}} \frac{1}{2} m_0^2 \hat\phi(\mathbf{x}_i)^2\ket{\Omega_{\phi(\mathbf{x}_i)}}.
 \label{eqn:scalar_E0diff_simple}
\end{equation}
Thus, we can obtain an upper bound on $ a   \left(E_0 - E'_{0,\phi(\mathbf{x_i})}\right) $ by evaluating the expectation value of the operator $\frac{1}{2} m_0^2 \hat\phi(\mathbf{x}_i)^2$ in the path integral
\begin{equation}
    \bra{\Omega_{\phi(\mathbf{x}_i)}} \frac{1}{2} m_0^2 \hat\phi(\mathbf{x}_i)^2\ket{\Omega_{\phi(\mathbf{x}_i)}} = \frac{\int  \mathcal{D}\phi  \frac{1}{2} m_0^2 \phi(\mathbf{x}_i)^2 e^{-S'_{\phi(\mathbf{x_i})}[\phi]}}{\int  \mathcal{D}\phi  e^{-S'_{\phi(\mathbf{x_i})}[\phi]}},
\end{equation}
where the modified action $S'_{\phi(\mathbf{x_i})}[\phi]$ is defined as
\begin{equation}
    S'_{\phi(\mathbf{x_i})}[\phi] = S[\phi] - \sum_t\frac{a_0}{a} \frac{1}{2} m_0^2 \phi(t, \mathbf{x}_i)^2.
    \label{eqn:scalar_S_phi_x_i}
\end{equation}
Here, $S[\phi]$ is the action corresponding to the original Hamiltonian $\hat{H}$ and is given by
\begin{equation}
 S[\phi] = \sum_{t, \mathbf{x}} \left[ \frac{a}{a_0}
\frac{(\phi (t + a_0, \mathbf{x}) - \phi (t,
\mathbf{x}))^2}{2} + \frac{a_0}{a} \frac{1}{2}
(\mathbf{\nabla} \phi)^2 (t, \mathbf{x}) +
\frac{a_0}{a} \frac{m_0^2 \phi (t, \mathbf{x})^2}{2} +
\frac{a_0}{a} \frac{\lambda_0}{4!} \phi (t, \mathbf{x})^4 
\right],
\label{eqn:scalar_S_unmodified}
\end{equation}
where $a_0$ stands for the lattice spacing in temporal direction, whereas $a$ is the lattice spacing in spatial direction.

In short, one can use the following workflow. First, either by calculating the energies directly or by evaluating $\bra{\Omega_{\phi(\mathbf{x}_i)}} \hat\phi(\mathbf{x}_i)^2\ket{\Omega_{\phi(\mathbf{x}_i)}}$, one can bound the energy difference $E_0 - E'_{0,\phi(\mathbf{x_i})}$. Plugging this value into Eqn.~\ref{eqn:scalar_bound_1}, one can obtain an upper bound for $\bra{\psi}\hat\phi(\mathbf{x}_i)^2\ket{\psi}$. Feeding this value into Eqn. \ref{eqn:jlp-phi-scaling}, the required cutoff $\phi_{\text{max}}$ can be obtained. In comparison, the original method in Ref.~\cite{jordan_quantum_2012} bounds $\bra{\psi}\hat\phi(\mathbf{x}_i)^2\ket{\psi}$ using Eqn.~\ref{eqn:jlp-energy} before feeding it into Eqn. \ref{eqn:jlp-phi-scaling} to obtain $\phi_{\text{max}}$. In their approach, $\bra{\psi}\hat\phi(\mathbf{x}_i)^2\ket{\psi}$ scales as $E_0+\Delta E$, whereas it scales as $E_0 - E'_{0,\phi(\mathbf{x_i})}+\Delta E$ in our method. Typically, the energy $E_0+\Delta E$ is dominated by $E_0$, which scales as volume. By contrast, $E_0 - E'_{0,\phi(\mathbf{x_i})}$ is usually much smaller. 
Hence, our method can considerably tighten the bound for $\bra{\psi}\hat\phi(\mathbf{x}_i)^2\ket{\psi}$, yielding a reduction in  $\phi_{\text{max}}$.

We note that, strictly speaking, obtaining reliable estimates for quantum field theories requires accounting for discretization effects and renormalization~\cite{carena_lattice_2021,kane_obtaining_2025}. While such an endeavor is beyond the scope of our paper, we offer two brief comments on this matter. First, the Hamiltonian generally needs to be renormalized. Here, we assume that spatial and temporal scale setting has already been performed and that the Hamiltonian parameters have been properly tuned. Given such a Hamiltonian, one can then construct the corresponding Lagrangian to perform classical Monte Carlo calculations (see Appendix~\ref{app:path_integral}). Second, at a given spatial lattice spacing, matching the quantities calculated from the Euclidean path integral to the Hamiltonian formulation requires taking the Hamiltonian limit of the path integral; that is, the temporal lattice spacing must be taken to zero. In practice, careful extrapolation to the $a_0\rightarrow 0$ limit is required.

In this section, we have described a technique to tighten the JLP bound with the help of Euclidean path-integral Monte Carlo methods. However, to further tighten the bound, we need to combine this with a second technique, which is discussed in the following section.

\subsubsection{Further improvement: incorporating ``$p$-norm trick''\label{sec:method:scalar:improved}}
Sec. \ref{sec:method:scalar:basic} lays out a basic method to tighten the upper bound for $\bra{\psi} \hat \phi(\mathbf{x}_i)^2 \ket{\psi}$. Nevertheless, recall from Eqn.~\ref{eqn:jlp-phi-scaling} that the cutoff $\phi_\text{max}$ needs to scale as $\sqrt{\mathcal{V}\bra{\psi}\hat\phi^2(\mathbf{x})\ket{\psi}/ \epsilon}$, where $\mathcal{V}$ is the total number of sites. Thus, even after tightening the bound    for $\bra{\psi} \hat\phi(\mathbf{x}_i)^2 \ket{\psi}$, there is still a $\sqrt{\mathcal{V}}$ factor in the cutoff. However, in certain cases, this limitation can be circumvented. To achieve this, we will use the ``$p$-norm trick.'' Instead of Eqn.~\ref{eqn:jlp-phi-scaling}, we will re-derive an expression for the cutoff $\phi_\text{max}$ using this trick, which enables further improvement in the bound. We will again use the scalar field theory as an example.

Instead of using Monte Carlo methods to bound $\bra{\psi} \hat\phi(\mathbf{x}_i)^2 \ket{\psi}$ for $\phi$ field at a single site $\mathbf{x}_i$, we will attempt to bound $\phi$ in a more ``global'' fashion. Let's define the ``$p$-norm'' for the $\phi$ field operator as
\begin{equation}
     \hat\phi^{(p)} = \left(\sum_{i=1}^{\mathcal{V}} |\hat\phi(\mathbf{x}_i)|^p\right)^{1/p}.
\end{equation}
When it acts on an eigenstate of the $\phi$ operator, we get
\begin{equation}
    \hat\phi^{(p)}  \ket{\phi_1, \phi_2, ..., \phi_\mathcal{V}} 
     = \left(|\phi_1|^p+|\phi_2|^p + ...+|\phi_\mathcal{V}|^p\right)^{1/p} \ket{\phi_1, \phi_2, ..., \phi_\mathcal{V}}.
     % \equiv \|\boldsymbol{\phi}\|  \ket{\phi_1, \phi_2, ..., \phi_\mathcal{V}},
\end{equation}
In particular, when $p=\infty$, we get the infinity-norm
\begin{equation}
     \hat\phi^{(\infty)}  \ket{\phi_1, \phi_2, ..., \phi_\mathcal{V}} 
     = \max\left(|\phi_1|,|\phi_2|, ...,|\phi_\mathcal{V}|\right)\; \ket{\phi_1, \phi_2, ..., \phi_\mathcal{V}}.
\end{equation}
For convenience, let's denote
\begin{equation}
    \| \boldsymbol{\phi}\|_p \equiv \left(|\phi_1|^p+|\phi_2|^p + ...+|\phi_\mathcal{V}|^p\right)^{1/p},
\end{equation}
so that we have $  \hat\phi^{(p)}  \ket{\phi_1,  ..., \phi_\mathcal{V}}=\| \boldsymbol\phi\|_p \ket{\phi_1,  ..., \phi_\mathcal{V}} $.

By analyzing  $ \| \boldsymbol{\phi}\|_p$ instead of $|\phi_i|$ at each single site, one can obtain an improved error bound compared to Eqn.~\ref{eqn:jlp_err_tot}. Recall from Eqn. \ref{eqn:state_in_phi_basis} that, for a given state $\ket{\psi}$, there is an associated probability distribution of $\phi_1,\phi_2,..., \phi_\mathcal{V}$, and the truncation error $\epsilon$ is characterized by the probability for $|\phi_i|$ at any site $i$ to exceed $\phi_\text{max}$. Ref.~\cite{jordan_quantum_2012} bounds this probability per site and then applies the union bound to bound the total error $\epsilon$,  thus introducing a $\sqrt{\mathcal{V}}$ factor. By contrast, in our approach, since $ \| \boldsymbol{\phi}\|_p\geq |\phi_i|$ for all $i$, the truncation error is upper bounded by the probability for $\| \boldsymbol{\phi}\|_p$ to exceed $\phi_\text{max}$. 
Note, this result should hold for general $p$. Here, we set $p$ to $\infty$. 

Similar to the single-site case, we can bound the probability for  $\| \boldsymbol{\phi}\|_\infty$ to exceed $\phi_\text{max}$  for a given state $\ket{\psi}$ by bounding $\sqrt{\bra{\psi}   (\hat\phi^{(\infty)} )^2 \ket{\psi}} $, which bounds both the mean value and the standard deviation for $\| \boldsymbol{\phi}\|_\infty$. By Chebyshev's inequality, we have 
\begin{equation}
    \epsilon \leq \left( \frac{\sqrt{\bra{\psi}   (\hat\phi^{(\infty)} )^2 \ket{\psi}} }{\phi_\text{max}-\sqrt{\bra{\psi}   (\hat\phi^{(\infty)} )^2 \ket{\psi}} }\right)^2,
\end{equation}
or in other words, given an error budget $\epsilon$, $\phi_\text{max}$ needs to be chosen to be
\begin{equation}
    \phi_\text{max} =
    % \sqrt{\frac{\bra{\psi}   (\hat\phi^{(p)} )^2 \ket{\psi}}{\epsilon}} + \sqrt{\bra{\psi}   (\hat\phi^{(p)} )^2 \ket{\psi}} 
   \left(\frac{1}{\sqrt{\epsilon}}+1\right) \sqrt{\bra{\psi}   (\hat\phi^{(\infty)} )^2 \ket{\psi}} 
    \sim \sqrt{\frac{\bra{\psi}   (\hat\phi^{(\infty)} )^2 \ket{\psi}}{\epsilon}} ,\label{eqn:scalar_cutoff_improved}
\end{equation}
which does not have the explicit $\sqrt{\mathcal{V}}$ factor appearing in the JLP bound in Eqn.~\ref{eqn:jlp-phi-scaling}.
The implicit dependence of $\sqrt{\bra{\psi}   (\hat\phi^{(\infty)} )^2 \ket{\psi}}$ on $\mathcal{V}$, as we will show, is usually much more moderate.

The bound for $\bra{\psi}   (\hat\phi^{(\infty)} )^2 \ket{\psi}$ can be estimated using a method similar to Eqn. \ref{eqn:bound-improve-single-1}.  We can define 
\begin{equation}
    \hat H'_{\phi^{(\infty)},\eta} = \hat H - \eta \frac{1}{a} \frac{1}{2}m_0^2 (\hat \phi^{(\infty)})^2,
\end{equation}
where $\eta$ is a chosen constant.
To ensure $\hat H'_{\phi^{(\infty)},\eta} $ is lower bounded,  $\eta$ should not be too large. Here, setting $\eta \leq 1$ is sufficient to ensure  $\hat H'_{\phi^{(\infty)},\eta} $ is bounded below. Let's denote the minimal eigenvalue of $\hat H'_{\phi^{(\infty)},\eta}$ by $E'_{0,\phi^{(\infty)},\eta}$ and its ground state by $\ket{\Omega_{\phi^{(\infty)},\eta}}$. Suppose one is given that $\bra{\psi} \hat{H} \ket{\psi}\leq E = E_0+\Delta E$, we can now bound $\bra{\psi}   (\hat\phi^{(\infty)} )^2 \ket{\psi}$ by 
\begin{equation}
    \begin{aligned}
        \bra{\psi}   (\hat\phi^{(\infty)} )^2 \ket{\psi} =& \frac{2}{\eta m_0^2} \left(a \bra{\psi}\hat{H} \ket{\psi}-a \bra{\psi}\hat H'_{\phi^{(\infty)},\eta} \ket{\psi}\right) \\ 
        \leq & \frac{2}{\eta m_0^2}a \left(E_0-E'_{0,\phi^{(\infty)},\eta}+\Delta E\right),
    \end{aligned}
    \label{eqn:scalar_phi_inf_bound}
\end{equation}
where $\Delta E$ is the physical energy scale to be chosen, while $a \left(E_0-E'_{0,\phi^{(\infty)},\eta}\right)$ can be bounded similar to Eqn. \ref{eqn:scalar_E0diff_simple}:
\begin{equation}
    \frac{2}{\eta m_0^2} a \left(E_0-E'_{0,\phi^{(\infty)},\eta}\right) \leq \bra{\Omega_{\phi^{(\infty)},\eta}}(\hat \phi^{(\infty)})^2\ket{\Omega_{\phi^{(\infty)},\eta}},\label{eqn:scalar_E0diff_improved}
\end{equation}
where the expectation value $\bra{\Omega_{\phi^{(\infty)},\eta}} (\hat \phi^{(\infty)})^2\ket{\Omega_{\phi^{(\infty)},\eta}}$ is accessible by Euclidean Monte-Carlo calculation:
\begin{equation}
    \bra{\Omega_{\phi^{(\infty)},\eta}}(\hat \phi^{(\infty)})^2\ket{\Omega_{\phi^{(\infty)},\eta}} =
   \frac{\int  \mathcal{D}\phi\left(\phi^{(\infty)}\right)^2 e^{-S'_{\phi^{(\infty)},\eta}[\phi]}}{\int  \mathcal{D}\phi  e^{-S'_{\phi^{(\infty)},\eta}[\phi]}},\label{eqn:scalar_phi_inf_path_integral}
\end{equation}
where the modified action is given by
\begin{equation}
    S'_{\phi^{(\infty)},\eta}[\phi] = S[\phi] - \sum_t\frac{a_0}{a} \eta \frac{1}{2} m_0^2 \left(\phi^{(\infty)}(t)\right)^2.\label{eqn:scalar_S_phimax}
\end{equation}

Putting Eqns. \ref{eqn:scalar_phi_inf_bound}, \ref{eqn:scalar_E0diff_improved}, and \ref{eqn:scalar_phi_inf_path_integral} together, we obtain
\begin{equation}
\begin{aligned}
    \bra{\psi}   (\hat\phi^{(\infty)} )^2 \ket{\psi} \leq  &\bra{\Omega_{\phi^{(\infty)},\eta}}(\hat \phi^{(\infty)})^2\ket{\Omega_{\phi^{(\infty)},\eta}} + \frac{2}{\eta m_0^2}a  \Delta E \\
    = &   \frac{\int  \mathcal{D}\phi\left(\phi^{(\infty)}\right)^2 e^{-S'_{\phi^{(\infty)},\eta}[\phi]}}{\int  \mathcal{D}\phi  e^{-S'_{\phi^{(\infty)},\eta}[\phi]}}  + \frac{2}{\eta m_0^2}a  \Delta E.
 \end{aligned}
 \label{eqn:scalar_phi_inf_bound_summary}
\end{equation}

To summarize, we propose the following procedure. One can use Monte Carlo methods to numerically evaluate $\bra{\Omega_{\phi^{(\infty)},\eta}}(\hat \phi^{(\infty)})^2\ket{\Omega_{\phi^{(\infty)},\eta}}$, which can be used by Eqn.~\ref{eqn:scalar_phi_inf_bound_summary} to bound $\bra{\psi}   (\hat\phi^{(\infty)} )^2 \ket{\psi}$. Having bounded $\bra{\psi}   (\hat\phi^{(\infty)} )^2 \ket{\psi}$, the required cutoff $\phi_\text{max}$ can then be obtained from Eqn.~\ref{eqn:scalar_cutoff_improved}.
Compared to the original JLP bound (see Eqn. \ref{eqn:jlp-phi-scaling}), this new approach improves the bound in two aspects. Firstly, Eqn. \ref{eqn:jlp-phi-scaling} contains an explicit $\sqrt{\mathcal{V}}$ factor, which is absent from Eqn. \ref{eqn:scalar_cutoff_improved}. Secondly, in the original JLP bound, there is another implicit $\sqrt{\mathcal{V}}$ factor in $\sqrt{\bra{\psi}\hat \phi^2(\mathbf{x})\ket{\psi}}$, since the inequality $\bra{\psi}\hat \phi^2(\mathbf{x})\ket{\psi}\leq \frac{2}{m_0^2} a E$ is used, where $E=E_0+\Delta E$ in general scales linearly with volume. By contrast, the expectation value $\bra{\psi}   (\hat\phi^{(\infty)} )^2 \ket{\psi}$ in this approach is bounded by Eqn. \ref{eqn:scalar_phi_inf_bound_summary}, which is not expected to exhibit such scaling. While $\bra{\psi}   (\hat\phi^{(\infty)} )^2 \ket{\psi}$ will still increase as the volume $\mathcal{V}$ increases, the dependence will typically be much milder than the linear dependence, as we will show in the numerical results in Sec. \ref{sec:results:phi}. 
In this way, one can greatly reduce the required truncation cutoff $\phi_\text{max}$, and the achieved saving factor, as will be shown, is close to $\mathcal{V}$ in practice.

\subsection{2+1D U(1) gauge theory dual formalism\label{sec:method:dual}}
The approach presented in the previous sections can be applied to other theories as well. A notable example is the dual formalism of (2+1)-dimensional U(1) gauge theory  \cite{bender_gauge_2020,bauer_efficient_2021,kane_efficient_2022}.

For the pure gauge theory, the degrees of freedom in this theory are the electromagnetic field residing on the plaquettes. For periodic boundary condition, there can also be two global loops besides the plaquettes. In this work, we focus on the open boundary condition. For each plaquette, there is a pair of conjugate fields: $\hat B$ and $\hat R$ such that $[\hat B_p, \hat R_q] = i \delta_{p,q}$, where $p$ and $q$ are indices denoting the location of the plaquettes and run from $1$ to $\mathcal{V}$, where $\mathcal{V}$ is the total number of plaquettes. $\hat{B}$ is the magnetic field operator, while $\hat{R}$ is the electric rotor operator. The Hamiltonian takes the form
\begin{equation}
    \hat H_{\text{U(1)}} = \frac{1}{a} \frac{g^2}{2} \sum_{p,q} \hat R_{p} M_{p,q} \hat R_{q}+\frac{1}{a}\frac{1}{2 g^2} \sum_{p} \left(\hat B_{p}\right)^2,
    \label{eqn:dual_hamiltonian}
\end{equation}
where the coupling $g$ has been rescaled to make it dimensionless. $M_{p,q}$ are the coefficients for the rotator  field term associated with square of the lattice curl; when $p$ and $q$ are interpreted as matrix indices, $M_{p,q}$ can be viewed as a real, symmetric, positive-definite matrix.
The magnetic term in an alternative formulation can be written as $-\frac{1}{a}\frac{1}{g^2}\sum_{p}  \cos\left(B_{p}\right)$, referred to as the compact formulation, while the expression in Eqn. \ref{eqn:dual_hamiltonian} is that of the non-compact formulation~\cite{bauer_efficient_2021}.

The $B$ and $R$ fields here play a similar role as the $\phi$ and $\pi$ fields in the scalar theory and likewise need to be truncated for practical quantum simulation. In this section, we discuss how the truncation error of $B$ field can be bounded by a tightened version of the energy-based bound, similar to the previous section.

As in the previous section, one can define $\hat B^{(\infty)}$ so that
\begin{equation}
    \hat B^{(\infty)} \ket{B_1, ..., B_\mathcal{V}} = \max(|B_1|, ..., |B_\mathcal{V}|)  \ket{B_1, ..., B_\mathcal{V}} .
\end{equation}
Similar to Eqn. \ref{eqn:scalar_cutoff_improved}, given an error budget $\epsilon$, the cutoff $B_\text{max}$ needs to scale as
\begin{equation}
    B_\text{max} = \left(\frac{1}{\sqrt{\epsilon}}+1\right) \sqrt{\bra{\psi} ( \hat B^{(\infty)} )^2 \ket{\psi}} \sim \sqrt{\frac{\bra{\psi} ( \hat B^{(\infty)} )^2 \ket{\psi}}{\epsilon}}.
    \label{eqn:dual_cutoff_improved}
\end{equation}
The expectation value $\bra{\psi} ( \hat B^{(\infty)} )^2 \ket{\psi}$ can be bounded using the energy-based bound assisted by Monte Carlo methods. Suppose that $\bra{\psi} \hat H_{\text{U(1)}} \ket{\psi} \leq E = E_0+\Delta E $, where $E_0$ is the ground state energy of $\hat H_{\text{U(1)}}$. One can define a modified Hamiltonian $\hat H'_{B^{(\infty)},\eta}$ as
\begin{equation}
    \hat H'_{B^{(\infty)},\eta} =  \hat H_{\text{U(1)}}  - \eta \frac{1}{a}\frac{1}{2 g^2} ( \hat B^{(\infty)} )^2 ,
\end{equation}
and denote its ground state as $\ket{\Omega_{B^{(\infty)},\eta}}$. Then, similar to Eqn. \ref{eqn:scalar_phi_inf_bound_summary}, we have
\begin{equation}
    \begin{aligned}
        \bra{\psi} ( \hat B^{(\infty)} )^2 \ket{\psi} \leq & \bra{\Omega_{B^{(\infty)},\eta}} ( \hat B^{(\infty)} )^2 \ket{\Omega_{B^{(\infty)},\eta}} + \frac{2 g^2}{\eta} a \Delta E \\
        =& \frac{\int \mathcal{D} B \; ( B^{(\infty)} )^2 e^{-S'_{B^{(\infty)},\eta}[B]}}{\int \mathcal{D} B  \; e^{-S'_{B^{(\infty)},\eta}[B]}}+ \frac{2 g^2}{\eta} a \Delta E,
    \end{aligned}
    \label{eqn:dual_inf_bound_summary}
\end{equation}
where the modified action is
\begin{equation}
    S'_{B^{(\infty)},\eta}[B] = \sum_t \left(\frac{a}{a_0} \frac{1}{2g^2}\sum_i \frac{\left( \sum_p U^T_{i p}\left(B_p(t+a_0) -B_p(t)\right)\right)^2}{\lambda_i}+\frac{a_0}{a}\frac{1}{2g^2}\sum_p (B_p)^2  - \eta \frac{a_0}{a}\frac{1}{2g^2} ( B^{(\infty)} )^2\right),
    \label{eqn:dual_S_Bmax}
\end{equation}
where the real positive values $\lambda_i$ and the orthonormal real matrix $U$ are defined by the diagonalization of the $M$ matrix: $M_{p,q} =\sum_i U_{p,i} \lambda_i U^\dagger_{i,q}$ (see Appendix~\ref{app:dual} for more information on $M_{p,q}$ matrix).

Using Eqn. \ref{eqn:dual_inf_bound_summary} combined with Eqn. \ref{eqn:dual_cutoff_improved}, an improved estimate for the cutoff $B_{\text{max}}$ can thus be obtained.  The numerical results can be found in Section~\ref{sec:results:B}.

\section{Methodology II: bounding the conjugate field\label{sec:methodII}}
In this section, we discuss methods to tighten the bounds for the conjugate field variable, i.e., $\pi$ field in scalar field theory and $R$ in the 2+1D U(1) gauge theory in dual formalism. Improving bounds for these variables turn out be less straight-forward than the $\phi$ field or the $B$ field, and the achievable resource saving is smaller. Nonetheless, one can still tighten the bound compared to the original JLP method. One may still use the ``Monte Carlo trick'' or the ``$p$-norm trick'' with some modifications.
The details are presented in the following subsections.

\subsection{$\pi$ field in $\phi^4$ theory\label{sec:methodII:pi}}
To bound the $\pi$ field, one cannot use a similar method as Section. \ref{sec:method:scalar:improved}. This is due to the fact that, unlike the case for $(\hat{\phi}^{(\infty)})^2$ operator, it is no longer straight-forward to evaluate $(\hat{\pi}^{(\infty)})^2$ operator in a path integral or defining a modified action based on the operator.
The path integral used in the Monte Carlo trick is obtained by integrating out the $\pi$ field (see Appendix~\ref{app:path_integral}). This step can be readily carried out via a Gaussian integral for the standard Hamiltonian, whose dependence on the $\pi$ field is quadratic. It also allows simple polynomial insertions, such as $\hat{\pi}(\mathbf{x})^2$, to be evaluated through Gaussian moments. However, a term involving $(\hat{\pi}^{(\infty)})^2$ does not reduce to the same elementary Gaussian calculation, making it unclear how to evaluate the relevant observable or obtain a modified action analogous to the $(\hat{\phi}^{(\infty)})^2$ case.

However, one can still derive an improved bound for the $\pi$ field. One method to do so is as follows. Instead of setting $p=\infty$ in the $p$-norm, one can use $p=2$ instead.
Namely, 
\begin{equation}
    \hat{\pi}^{(2)} \equiv \sqrt{\sum_{\mathbf{x}_i} ( \hat{\pi}_{\mathbf{x}_i})^2}.
\end{equation}

Similar to Eqn. \ref{eqn:state_in_phi_basis}, one can expand a state $\ket{\psi}$ in the $\pi$ basis \cite{jordan_quantum_2012}, giving a probability distribution in $\pi(\mathbf{x}_1)$, $\pi(\mathbf{x}_2)$, ..., $\pi(\mathbf{x}_\mathcal{V})$. By a similar argument, for a chosen truncation $\pi_{\text{max}}$, the truncation error $\epsilon$ is the probability that any of $|\pi(\mathbf{x}_1)|$ ,$|\pi(\mathbf{x}_2)|$, $|\pi(\mathbf{x}_\mathcal{V})|$ exceeds $\pi_{\text{max}}$. Defining $\|\boldsymbol{\pi}\|_2=\sqrt{\sum_{\mathbf{x}_i} ( {\pi}_{\mathbf{x}_i})^2}$, one can see that $\|\boldsymbol{\pi}\|_2$ upper bounds all of $|\pi(\mathbf{x}_1)|$ ,$|\pi(\mathbf{x}_2)|$, $|\pi(\mathbf{x}_\mathcal{V})|$. Thus, the truncation error is upper bounded by the probability for $\|\boldsymbol{\pi}\|_2$ to exceed $\pi_{\text{max}}$. Thus, similar to the argument in Eqn. \ref{eqn:scalar_cutoff_improved}, one can choose the required $\pi_{\text{max}}$ for a given $\epsilon$ as
\begin{equation}
    \pi_{\text{max}} = 
      \left(\frac{1}{\sqrt{\epsilon}}+1\right) \sqrt{\bra{\psi}   (\hat\pi^{(2)} )^2 \ket{\psi}} 
    \sim \sqrt{\frac{\bra{\psi}   (\hat\pi^{(2)} )^2\ket{\psi}}{\epsilon}} 
    \leq \sqrt{\frac{ 2 a E}{\epsilon}},
    \label{eqn:scalar_pi_cutoff_im}
\end{equation}
where we have used that $\frac{1}{a}\bra{\psi}  \frac{1}{2} (\hat\pi^{(2)} )^2\ket{\psi}=\frac{1}{a} \bra{\psi}  \frac{1}{2} \sum_{\mathbf{x}_i} ( \hat{\pi}_{\mathbf{x}_i})^2\ket{\psi}\leq E$ assuming $\bra{\psi} \hat{H} \ket{\psi}\leq E$.

Notice, in general, unlike $\langle \frac{1}{2}(\pi^{(\infty)})^2 \rangle$, which should have a very mild dependence on volume, the expectation value $\bra{\psi}  \frac{1}{2} (\hat\pi^{(2)} )^2\ket{\psi}$ should scale linearly with volume because it is a sum of $\mathcal{V}$ positive terms. Hence, in terms of asymptotic scaling in $\mathcal{V}$, one cannot obtain a more optimized bound than $\frac{1}{a}\bra{\psi}  \frac{1}{2} (\hat\pi^{(2)} )^2\ket{\psi}\leq E$. 
For this reason, combining the Monte Carlo technique used in Section \ref{sec:method:scalar:improved} with the $p$-norm technique here would likely yield only marginal, if any, further benefit.

In comparison, using the original JLP bound \cite{jordan_quantum_2012}, one obtains
\begin{equation}
    \pi_{\text{max}} = 
      \left(\sqrt{\frac{\mathcal{V}}{\epsilon}}+1\right) \sqrt{\bra{\psi}   (\hat\pi(\mathbf{x}_i) )^2 \ket{\psi}} \sim \sqrt{\frac{\mathcal{V}\bra{\psi}   (\hat\pi(\mathbf{x}_i) )^2\ket{\psi}}{\epsilon}} 
    \leq \sqrt{\frac{\mathcal{V}\cdot 2 a E}{\epsilon}},
    \label{eqn:scalar_pi_cutoff_original}
\end{equation}
where $\mathbf{x}_i$ is one of the lattice sites and $\frac{1}{a}\bra{\psi}  \frac{1}{2} (\hat\pi(\mathbf{x}_i) )^2\ket{\psi}\leq E$ is used. 

Comparing Eqn. \ref{eqn:scalar_pi_cutoff_im} to Eqn. \ref{eqn:scalar_pi_cutoff_original}, our method tightens the $\pi_\text{max}$ cutoff by a factor of $\sqrt{\mathcal{V}}$.
Notice, while Eqn. \ref{eqn:scalar_pi_cutoff_im} is free of the explicit $\sqrt{\mathcal{V}}$ factor in Eqn. \ref{eqn:scalar_pi_cutoff_original}, there is still an implicit $\sqrt{\mathcal{V}}$ dependence from the $\sqrt{E}$ factor.
A factor $\sqrt{V}$ improvement is not as drastic as the nearly a factor of $\mathcal{V}$ improvement in the $\phi$ field case (see Sections \ref{sec:method:scalar:improved} and \ref{sec:results:phi}). 
Nevertheless, this is still a considerable improvement. Numerical results are presented in Section~\ref{sec:results:pi}.

While the $p$-norm trick is most powerful when combined with the Monte Carlo trick (see Section~\ref{sec:method:scalar:improved}), the present example shows that it can also be useful as a standalone technique in settings where a Monte Carlo calculation is unavailable or unnecessary. 
In particular, when the truncation cutoff can be effectively controlled by bounding the energy contribution of an aggregate term analogous to $\bra{\psi} \sum_{\mathbf{x}_i} (\hat \pi_{\mathbf{x}_i})^2 \ket{\psi}$ rather than individual terms, as in the scalar field theory considered here, one may obtain an analytic reduction of the field truncation by a factor of $\sqrt{\mathcal{V}}$ ``for free,'' with minimal additional effort required.

Note, instead of the $p$-norm trick here, one can alternatively use the Monte-Carlo trick similar to Section \ref{sec:method:scalar:basic} to tighten the bound for $\sqrt{\bra{\psi} (\hat\pi(\mathbf{x}_i) )^2 \ket{\psi}}$ at a single site, achieving a factor of $\sqrt{\mathcal{V}}$ saving. 
Though, the two tricks cannot be combined together to achieve further improvement.

\subsection{$R$ field\label{sec:methodII:R}}
For the electric rotor field $R$ in the dual formalism of 2+1D non-compact U(1) gauge theory, one can also obtain an improved cutoff. The method as in Section~\ref{sec:method:dual} will not be applicable due to the difficulty of evaluating the $\hat{R}^{(\infty)}$ field in a path integral or incorporating this term in the action. In the last section, we described a method where we work with $ \hat{\pi}^{(2)}$ instead of $ \hat{\pi}^{(\infty)}$. However, there are cases where this method would be ineffective. For example, in the 2+1D dual formalism U(1) gauge theory here, the maximum $\lambda_{\text{max}}$ such that $\hat{H}_{\text{U(1)}}-\lambda_{\text{max}}\frac{1}{a}\frac{g^2}{2} (\hat R^{(2)})^2$ stays positive-semidefinite will decrease considerably fast as the system volume decreases (see Appendix \ref{app:dual:eta}), which renders the bound $\bra{\psi} (\hat R^{(2)})^2\ket{\psi}\leq \frac{1}{\lambda_{\text{max}}} \frac{2}{g^2} a E$ looser and looser as the lattice size increases. 
Hence, in this section, we will outline an alternative method.

To bound the $R$ field, we will use the Monte Carlo trick for a single site, similar to Section \ref{sec:method:scalar:basic}. 
Rather than working with a global operator $\hat R^{(2)}$ or $\hat R^{(\infty)}$, we will bound the field value $R_p$ at each individual plaquette $p$.
Given an error budget $\epsilon$, by the union bound, it is sufficient to make sure the truncation error at each plaquette is smaller than $\epsilon/\mathcal{V}$. Hence, the field truncation value at plaquette $p$ 
\footnote{Because the open boundary condition breaks the translation symmetry in plaquettes, a cutoff $B_{p,\text{max}}$ needs to be determined for each plaquette $p$. This incurs a polynomial overhead in the classical Monte Carlo computation; however, this is acceptable given the substantially lower cost of classical resources compared to quantum computation.} 
can be chosen as
\begin{equation}
    R_{p,\text{max}} = \left(\sqrt{\frac{\mathcal{V}}{\epsilon}}+1\right) \sqrt{\bra{\psi}   (\hat R_p )^2 \ket{\psi}} \sim \sqrt{\frac{\mathcal{V}\bra{\psi}   (\hat R_p)^2\ket{\psi}}{\epsilon}}. 
    \label{eqn:Rp_cutoff}
\end{equation}
Then, instead of using the simple bound  $\bra{\psi}   (\hat R_p)^2\ket{\psi}\leq \frac{2}{g^2} a E$, we will use Monte Carlo methods to obtain an improved bound. One can choose a value $\eta$ such that $\sum_{r,s} \hat R_{r} M_{r,s} \hat R_{s} -\eta (\hat R_p)^2 $ is still positive-semidefinite, so that the modified Hamiltonian $\hat H'_{R_p, \eta}\equiv \hat H_{\text{U(1)}}-\eta\frac{1}{a}\frac{g^2}{2} (\hat R_p)^2$ will stay lower bounded. 
The maximal value for $\eta$ has a mild volume-dependence; see Appendix~\ref{app:dual:eta} for more details. 
We denote the ground state of the modified Hamiltonian by $\ket{\Omega_{R_p, \eta}}$. Denote $M'_{r,s}\equiv M_{r,s}-\eta \delta_{r,p} \delta_{p,s}$, and denote its diagonalization as $M'_{r,s} = \sum_i U'_{r,i} \lambda'_i U'^\dagger_{i,s}$.
Then, similar to Section \ref{sec:method:scalar}, assuming $\bra{\psi}  \hat H\ket{\psi}\leq E = E_0+\Delta E$, one can write
\begin{equation}
    \begin{aligned}
        \bra{\psi}   (\hat R_p)^2\ket{\psi} \leq & 
        \bra{\Omega_{R_p, \eta}}   (\hat R_p)^2\ket{\Omega_{R_p, \eta}}
        + \frac{2}{\eta g^2} a \Delta E \\
        =& \frac{\int \mathcal{D} B \; \tilde {R_p^2} e^{-S'_{R_p, \eta}[B]}}{\int \mathcal{D} B \; e^{-S'_{R_p, \eta}[B]}}
         + \frac{2}{\eta g^2} a \Delta E ,
    \end{aligned}
    \label{eqn:Rp_bound_summary}
\end{equation}
where the modified action $S'_{R_p}$ is defined as
\begin{equation}
    S'_{R_p, \eta} = \sum_t \left(\frac{a}{a_0} \frac{1}{2g^2}\sum_i \frac{\sum_q \left((U')^T_{i,q}\left(B_q(t+a_0) -B_q(t)\right)\right)^2}{\lambda'_i}+\frac{a_0}{a}\frac{1}{2g^2}\sum_q (B_q)^2  \right),
    \label{eqn:dual_R_action}
\end{equation}
and $\tilde {R_p^2} $ is an expression in terms of $B$ that corresponds to the expectation value of $\hat{R}_p^2$, defined as
\begin{equation}
    \tilde {R_p^2} = \sum_{i,j} U'_{p,i} U'_{p,j} \left( \frac{a \delta_{i,j}}{a_0 g^2 \lambda'_i}-\left(\frac{a \sum_q (U')^T_{i,q}(B_q(t+a_0) -B_q(t))}{a_0 g^2 \lambda'_i}\right)\left(\frac{a \sum_r (U')^T_{j,r}(B_r(t+a_0) -B_r(t))}{a_0 g^2 \lambda'_j}\right)\right),
    \label{eqn:dual_R_R2tilde}
\end{equation}
where the detailed derivation is given in Appendix~\ref{app:path_integral:dual}.

Hence, the workflow will be as follows. Using Monte Carlo methods, one can numerically estimate $\bra{\Omega_{R_p, \eta}}   (\hat R_p)^2\ket{\Omega_{R_p, \eta}}$. Then, from Eqn.~\ref{eqn:Rp_bound_summary}, one can obtain an upper bound for $\bra{\psi}   (\hat R_p)^2\ket{\psi}$. This leads to an estimate for the required cutoff $R_{p,\text{max}}$ by Eqn.~\ref{eqn:Rp_cutoff}. 
Compared to the simple bound where one uses $\bra{\psi}   (\hat R_p)^2\ket{\psi}\leq \frac{2}{g^2} a E$ with $E$ scaling linearly with system volume, the bound obtained in Eqn. \ref{eqn:Rp_bound_summary} will have little volume dependency. Hence, with this method, a factor of $\sqrt{\mathcal{V}}$ saving is expected. 
Thus, in cases where the $p$-norm technique used in Sec.~\ref{sec:methodII:pi} is unsuitable, one can still apply the Monte Carlo technique described here and achieve a similar level of resource saving.
The numerical results are shown in Section \ref{sec:results:R}.

\section{Numerical results\label{sec:results}}
In this section, we showcase the numerical results for the improved energy based bound.
The Euclidean path integral calculations were performed with a standard Metropolis algorithm.
After an initial thermalization phase, configurations were sampled every few hundred iterations, depending on the system volume.
To properly account for autocorrelation effects, the statistical uncertainties of the estimated observables were derived using a binning analysis.
The temporal lattice spacing is set to be $a_0=0.05$ in lattice units $a$.
For more details, we refer readers to the \href{https://github.com/yjh-bill/boson_truncation_study}{source code}.

\subsection{Bounding $\phi$ field in $\phi^4$ theory\label{sec:results:phi}}

\begin{figure}[h]
    \centering
    \includegraphics[width=0.9\linewidth]{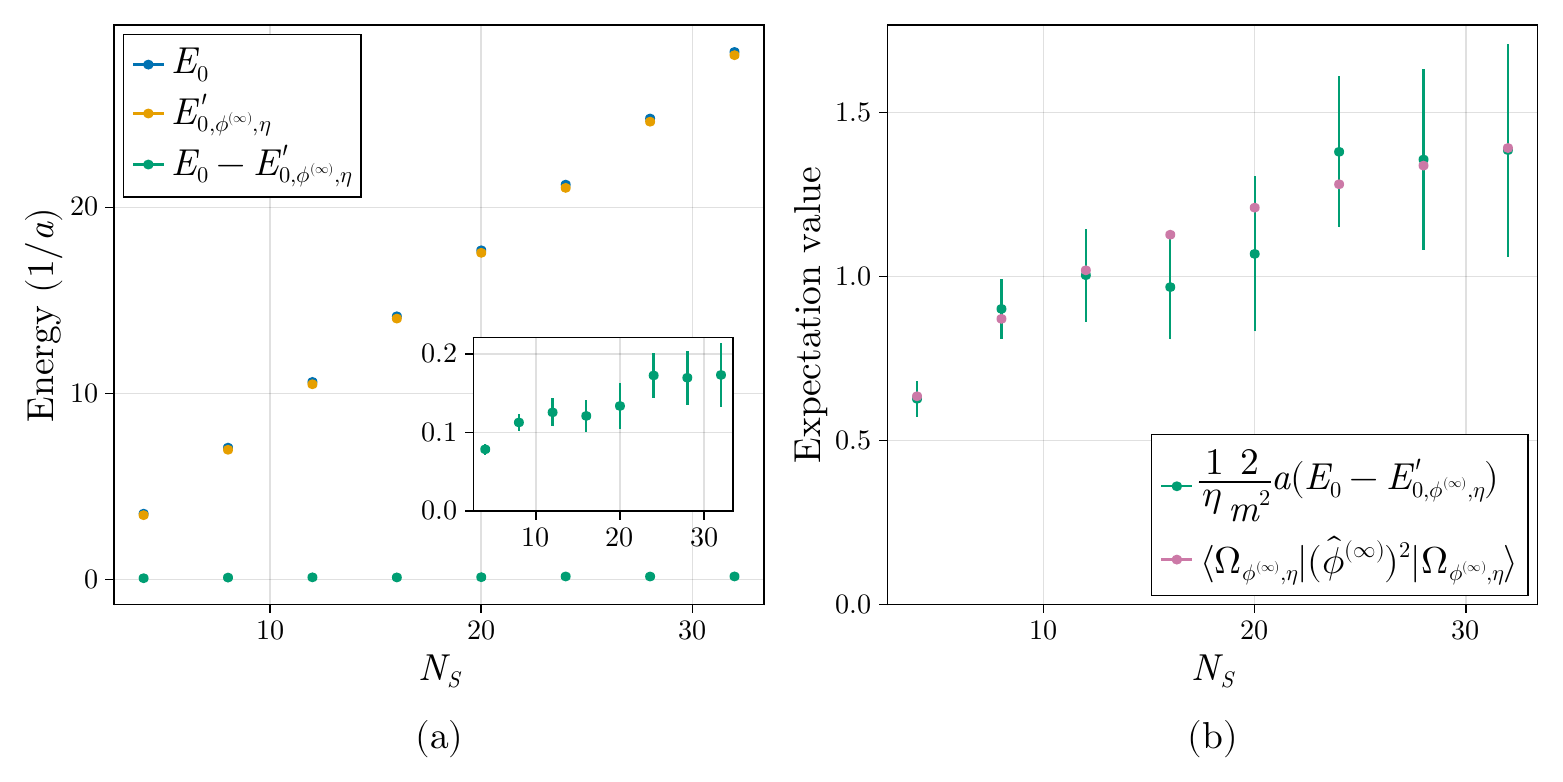}
    \caption{Comparison between energy scales. (a) the ground state energy $E_0$ of $\hat{H}$,  $E'_{0,\phi^{(\infty)},\eta}$ of $\hat H'_{\phi^{(\infty)},\eta}$, and their difference $E_0-E'_{0,\phi^{(\infty)},\eta}$ at different lattice sizes $N_S$.  The inset plot shows a zoomed-in plot of $E_0-E'_{0,\phi^{(\infty)},\eta}$.  (b) the value of $\frac{2}{\eta m_0^2} a \left(E_0-E'_{0,\phi^{(\infty)},\eta}\right)$ and $\bra{\Omega_{\phi^{(\infty)},\eta}}(\hat \phi^{(\infty)})^2\ket{\Omega_{\phi^{(\infty)},\eta}}$, either of which can be used to derive a bound for $\bra{\psi}   (\hat\phi^{(\infty)} )^2 \ket{\psi}$ (see Eqns. \ref{eqn:scalar_phi_inf_bound} and \ref{eqn:scalar_phi_inf_bound_summary} ).   
    }
    \label{fig:phi4_m0p5_lamb16_energy_plot}
\end{figure}

As described in Sec.~\ref{sec:method:scalar:improved}, to obtain the cutoff $\phi_{\text{max}}$ in the $\phi^4$ theory, one needs to first upper bound $\bra{\psi}   (\hat\phi^{(\infty)} )^2 \ket{\psi}$, which requires a numerical estimation for $\bra{\Omega_{\phi^{(\infty)},\eta}}(\hat \phi^{(\infty)})^2\ket{\Omega_{\phi^{(\infty)},\eta}}$. 
Hence, we perform a Monte Carlo calculation to evaluate it; at the same time, we also compute the ground state energies $E_0$ of $\hat{H}$ and $E'_{0,\phi^{(\infty)},\eta}$ of $\hat H'_{\phi^{(\infty)},\eta}$ to support our claim that  $E_0$ and $E'_{0,\phi^{(\infty)},\eta}$ scale linearly with volume, while their difference has a much weaker scaling. For this purpose, we use path integral Monte Carlo with anisotropic lattice spacing. 
In lattice units $a$, 
the bare mass is chosen to be $m_0 = 0.5$, and the bare coupling is chosen to be $\lambda_0=16$. The factor $\eta$ is set to $1$ here. 
The temporal extent of the lattice is 500 sites.
The results are shown in Fig.~\ref{fig:phi4_m0p5_lamb16_energy_plot}. In Fig. \ref{fig:phi4_m0p5_lamb16_energy_plot}(a), we plot $E_0$, $E'_{0,\phi^{(\infty)},\eta}$, and $E_0-E'_{0,\phi^{(\infty)},\eta}$ against the spatial lattice size $N_S$. As shown in the plot, $E_0$ and $E'_{0,\phi^{(\infty)},\eta}$ have roughly linear scaling as the volume, both being much larger than $1/a$, whereas their difference $E_0-E'_{0,\phi^{(\infty)},\eta}$ is much smaller at large volumes. 
Considering that the ``physical'' energy $\Delta E$ as in Eqn. \ref{eqn:scalar_bound_1} will typically be small compared to $1/a$ in scattering, the dominant source of energy in Eqn. \ref{eqn:scalar_bound_1} for the JLP bound will be the vacuum energy $E_0$ rather than $\Delta E$. 
In Fig.~\ref{fig:phi4_m0p5_lamb16_energy_plot}(b),  we plot the value of $\frac{2}{\eta m_0^2} a \left(E_0-E'_{0,\phi^{(\infty)},\eta}\right)$ and $\bra{\Omega_{\phi^{(\infty)},\eta}}(\hat \phi^{(\infty)})^2\ket{\Omega_{\phi^{(\infty)},\eta}}$ at different $N_S$, the latter of which is an upper bound for the former according to Eqn.~\ref{eqn:scalar_E0diff_improved}.
As shown in the figure, up to statistical uncertainty, $\bra{\Omega_{\phi^{(\infty)},\eta}}(\hat \phi^{(\infty)})^2\ket{\Omega_{\phi^{(\infty)},\eta}}$ provides a reasonably tight upper bound of $\frac{2}{\eta m_0^2} a \left(E_0-E'_{0,\phi^{(\infty)},\eta}\right)$, while having much less statistical uncertainty. 
Also, notably, as one can observe from the figure, $\bra{\Omega_{\phi^{(\infty)},\eta}}(\hat \phi^{(\infty)})^2\ket{\Omega_{\phi^{(\infty)},\eta}}$ does increase as the system volume increases, but the growth is very mild compared to the increase in $E_0$. 
Hence, in general, one can typically achieve a much tighter energy-based bound using the energy difference $E_0+\Delta E-E'_{0,\phi^{(\infty)},\eta}$ rather than the total energy $E_0 + \Delta E$

\begin{figure}[h]
    \centering
    \includegraphics[width=0.9\linewidth]{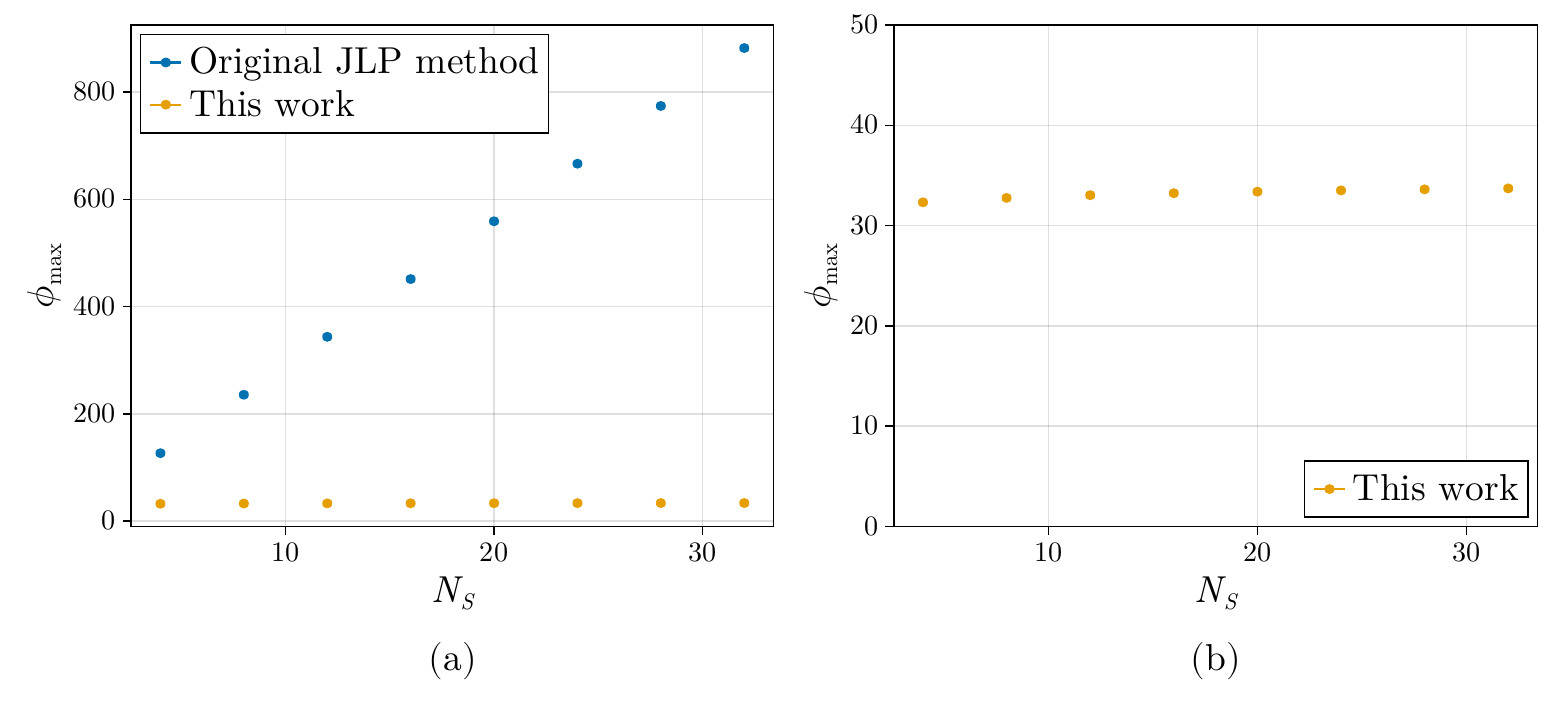}
    \caption{Estimation of the required truncation cutoff $\phi_\text{max}$ at different lattice sizes $N_S$.  
    The error budget is set to be $\epsilon=0.01$, and the energy above vacuum $\Delta E$ is set to $1/a$. (a) the comparison between the original JLP method and the approach in this work. (b) a zoomed-in view of the truncation scales of the latter.  }
    \label{fig:phi4_m0p5_lamb16_bound_compare}
\end{figure}

With $\bra{\Omega_{\phi^{(\infty)},\eta}}(\hat \phi^{(\infty)})^2\ket{\Omega_{\phi^{(\infty)},\eta}}$ estimated, one can obtain a bound for the truncation error.
For a given value of $\Delta E$, by putting the value of $\bra{\Omega_{\phi^{(\infty)},\eta}}(\hat \phi^{(\infty)})^2\ket{\Omega_{\phi^{(\infty)},\eta}}$ back into Eqn. \ref{eqn:scalar_phi_inf_bound_summary}, one can obtain a bound for $\bra{\psi}   (\hat\phi^{(\infty)} )^2 \ket{\psi}$ if $\bra{\psi} \hat{H}\ket{\psi} \leq E_0+\Delta E$. Then, plugging the bound for $\bra{\psi}   (\hat\phi^{(\infty)} )^2 \ket{\psi}$ into Eqn. \ref{eqn:scalar_cutoff_improved}, one can obtain an optimized bound for the truncation $\phi_{\text{max}}$ given an error budget $\epsilon$, as $\phi_\text{max}\geq 
   \left(\frac{1}{\sqrt{\epsilon}}+1\right) \sqrt{\bra{\psi}   (\hat\phi^{(\infty)} )^2 \ket{\psi}} $. In Fig.~\ref{fig:phi4_m0p5_lamb16_bound_compare} (a), we plot the truncation $\phi_\text{max}$ in our method as compared to the bound in the original JLP method. The error budget $\epsilon$ is set to $0.01$. The energy relative to the ground state is set to $\Delta E \leq 1/a$, which should be sufficiently large for scattering problems.
   In Fig. \ref{fig:phi4_m0p5_lamb16_bound_compare} (b), we present a zoomed-in plot of the truncation bounds in this approach.  As shown in the plot, at larger values of $N_S$, 
the improved estimate for $\phi_\text{max}$
is smaller than the original cutoff by a large factor. This demonstrates that the original energy-based bound can be significantly tightened.

To understand the asymptotic scaling of $\phi_\text{max}$ at large volumes, one can examine Eqns. \ref{eqn:scalar_phi_inf_bound_summary} and \ref{eqn:scalar_cutoff_improved} and note that the cutoff $\phi_\text{max}$ scales as $\sqrt{\bra{\Omega_{\phi^{(\infty)},\eta}}(\hat \phi^{(\infty)})^2\ket{\Omega_{\phi^{(\infty)},\eta}}}$ as the volume increases for a fixed $\Delta E$. 
The numerical results for the volume scaling of $\bra{\Omega_{\phi^{(\infty)},\eta}}(\hat \phi^{(\infty)})^2\ket{\Omega_{\phi^{(\infty)},\eta}}$ can be found in Fig.~\ref{fig:phi4_m0p5_lamb16_energy_plot}(b), but an analytical form is difficult to obtain. To gain physical intuition, consider the simplified scenario in which the field value $\phi(\mathbf{x}_i)$ at each site followed a Gaussian distribution independently, then the expectation value $\langle(\hat \phi^{(\infty)})^2\rangle$ could be shown to scale logarithmically at large volume \cite{gumbel_statistics_1958,arnold_first_2008}. Realistically, the distribution of $\phi$ inside the $\ket{\Omega_{\phi^{(\infty)},\eta}}$ state is much more complex and involves inter-site correlations; thus, the scaling may not remain logarithmic. Nevertheless, as shown in Fig.~\ref{fig:phi4_m0p5_lamb16_energy_plot}(b), the volume dependence of $\bra{\Omega_{\phi^{(\infty)},\eta}}(\hat \phi^{(\infty)})^2\ket{\Omega_{\phi^{(\infty)},\eta}}$ is very mild, appearing to be logarithmic or algebraic with a low exponent.

\subsection{Bounding $B$ field in 2+1D U(1) gauge theory in dual formalism\label{sec:results:B}}

\begin{figure}[h!]
    \centering
    \includegraphics[width=0.9\linewidth]{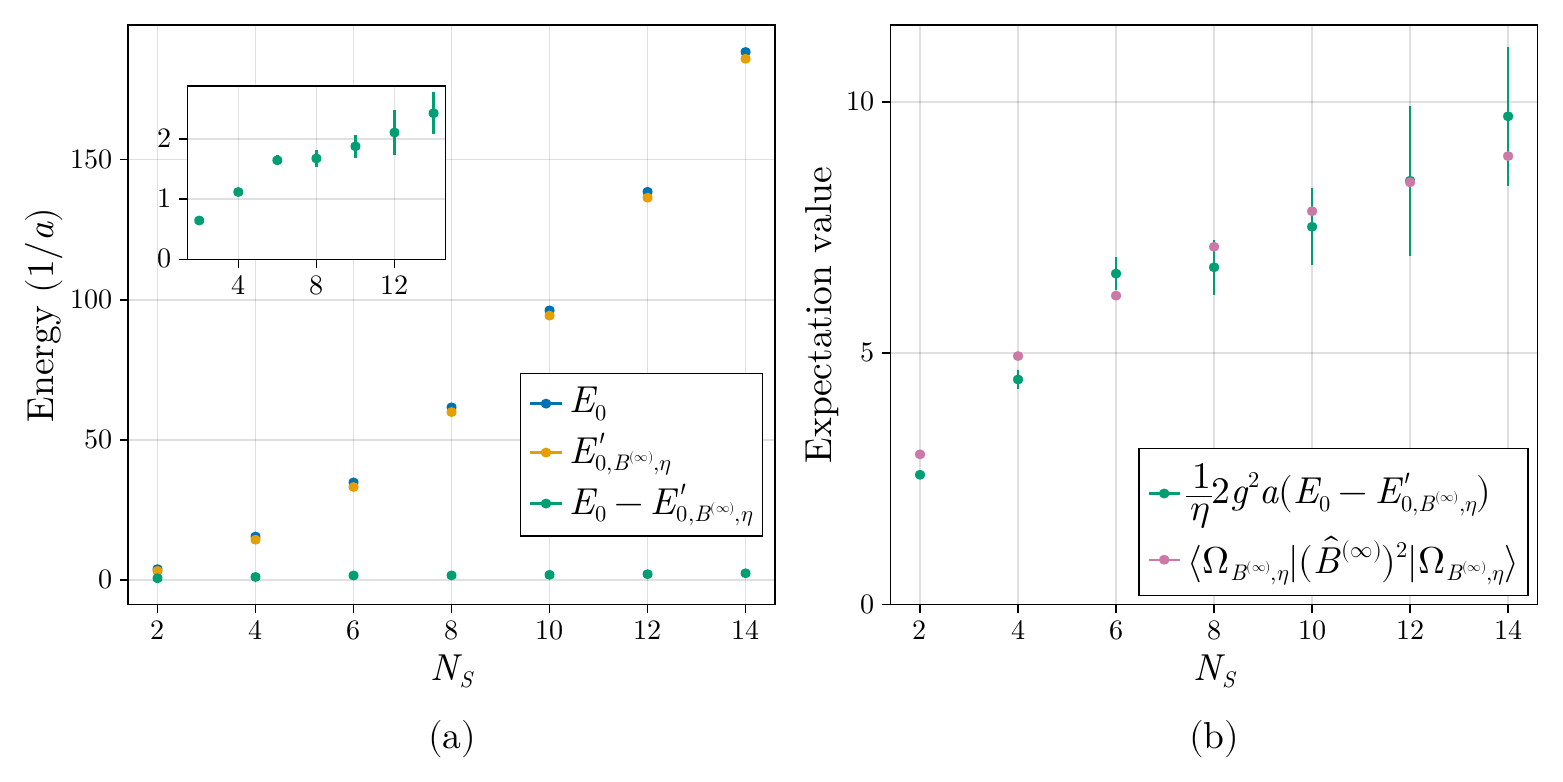}
    \caption{Comparison between energy scales. (a) the ground state energy $E_0$ of $\hat H_{\text{U(1)}}$,  $E'_{0,B^{(\infty)},\eta}$ of $\hat H'_{B^{(\infty)},\eta}$, and their difference $E_0-E'_{0,B^{(\infty)},\eta}$ at different lattice sizes $N_S$. Each lattice has $N_S\times N_S$ plaquettes spatially. 
    The inset plot shows a zoomed-in view of $E_0-E'_{0,B^{(\infty)},\eta}$.
    (b) $ \frac{1}{\eta} 2g^2 a (E_0-E'_{0,B^{(\infty)},\eta})$ and its upper bound $\bra{\Omega_{B^{(\infty)},\eta}} ( \hat B^{(\infty)} )^2 \ket{\Omega_{B^{(\infty)},\eta}}$. 
    }
    \label{fig:dual_g1_energy_plot}
\end{figure}

Similarly, we performed an anisotropic path-integral Monte Carlo simulation for the 2+1D U(1) gauge theory in dual formalism. 
In lattice unit $a$, the bare coupling is chosen to be $g=1.0$. 
The temporal extent of the lattice is 200 sites.
The factor $\eta$ is set to be $1/2$. The calculation is performed on $N_S\times N_S$ plaquettes with open boundary condition. Using the Monte Carlo calculation, we obtain the ground state energy $E_0$ of the Hamiltonian $\hat{H}_{\text{U(1)}}$, the ground state energy  $E'_{0,B^{(\infty)},\eta}$ of $\hat H'_{B^{(\infty)},\eta}$, and their difference $E_0-E'_{0,B^{(\infty)},\eta}$. The results are shown in Fig. \ref{fig:dual_g1_energy_plot}(a). In Fig. \ref{fig:dual_g1_energy_plot}(b), we plot $ \frac{1}{\eta} 2g^2 a (E_0-E'_{0,B^{(\infty)},\eta})$ along with its upper bound $ \bra{\Omega_{B^{(\infty)},\eta}} ( \hat B^{(\infty)} )^2 \ket{\Omega_{B^{(\infty)},\eta}}$. Similar to the $\phi^4$ theory case, $E_0$ and $E'_{0,B^{(\infty)},\eta}$ become large at large volumes. 
On the other hand, the volume dependence of $E_0-E'_{0,B^{(\infty)},\eta}$ or $\bra{\Omega_{B^{(\infty)},\eta}} ( \hat B^{(\infty)} )^2 \ket{\Omega_{B^{(\infty)},\eta}}$ is much milder, suggesting that they can offer a tighter energy-based bound at large volumes.

\begin{figure}[h!]
    \centering
    \includegraphics[width=0.9\linewidth]{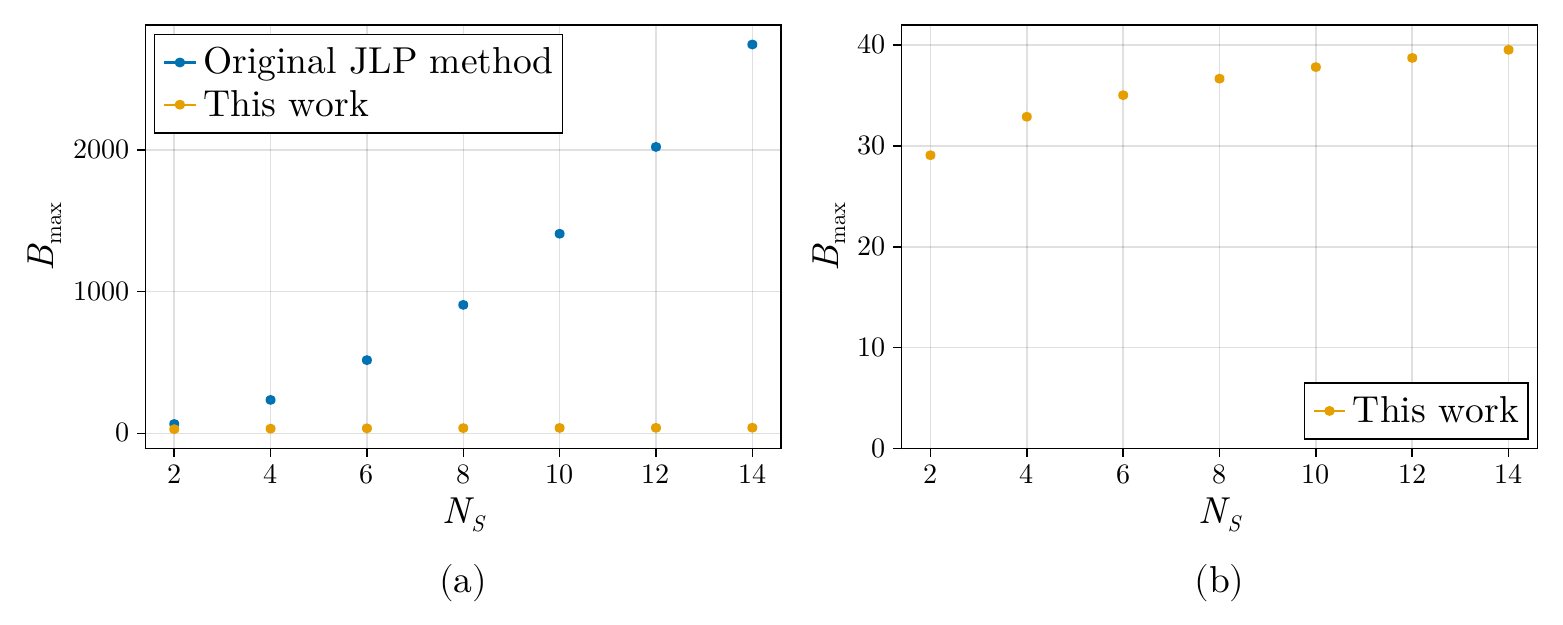}
    \caption{Estimation of the required truncation cutoff $B_\text{max}$ at different lattice size. Each lattice has $N_S\times N_S$ plaquettes spatially.   
   The error budget $\epsilon$ is set to $0.01$; the allowed energy above vacuum $\Delta E$ is set to $1/a$. (a) the comparison  between the original energy-based bound and the approach in this work. (b) a zoomed-in view of the truncation scales of the latter. }
    \label{fig:dual_g1_bound_compare}
\end{figure}

Once $ \bra{\Omega_{B^{(\infty)},\eta}} ( \hat B^{(\infty)} )^2 \ket{\Omega_{B^{(\infty)},\eta}}$ has been bounded, one can use Eqns. \ref{eqn:dual_cutoff_improved} and \ref{eqn:dual_inf_bound_summary} to obtain the needed cutoff $B_\text{max}$ given an error budget $\epsilon$. In Fig. \ref{fig:dual_g1_bound_compare}, we compare the $B_\text{max}$ as obtained in our approach with the $B_\text{max}$ obtained from the original JLP method. 
The error budget $\epsilon$ is set to $0.01$, and the allowed energy with respect to vacuum $\Delta E$ is set to be $1/a$.
As shown in the figure, the improved bound is significantly tighter than the original energy-based bound.

\subsection{Bounding $\pi$ field in $\phi^4$ theory\label{sec:results:pi}}
\begin{figure}[h!]
    \centering
    \includegraphics[width=0.9\linewidth]{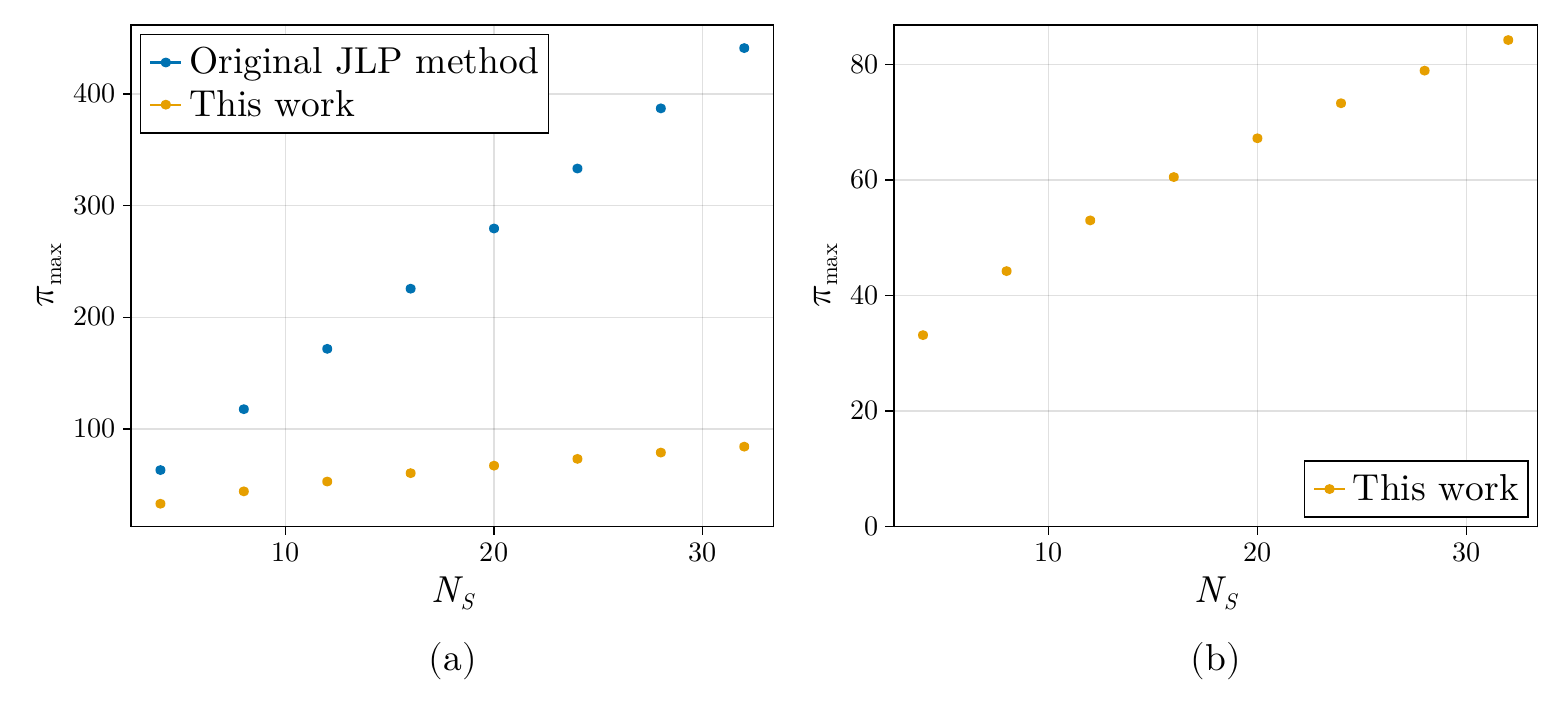}
    \caption{Estimation of the required truncation cutoff for $\pi$ field at different lattice sizes.  The error budget $\epsilon$ is set to $0.01$. $a \Delta E$ is set to $1$. (a) comparison of the obtained $\pi_\text{max}$  between the original JLP method and the approach in this work. (b) a zoomed-in view of the latter.  }
    \label{fig:phi4_m0p5_lamb16_pi_bound_compare}
\end{figure}

Using the method in Section~\ref{sec:methodII:pi}, we can obtain a bound for truncation error in $\pi$ field. We assume $\bra{\psi} \hat H \ket{\psi}\leq E = E_0+\Delta E$, where $E_0$ is the vacuum energy of $\hat H$. As a numerical example, we set $a \Delta E=1$, while $E_0$ is estimated using path integral Monte Carlo method. 
The settings are similar to those in Section~\ref{sec:results:phi}.
According to our improved bound Eqn. \ref{eqn:scalar_pi_cutoff_im}, the cutoff for $\pi$ should be chosen as $\pi_\text{max}\sim\sqrt{\frac{ 2 a E}{\epsilon}}$; whereas by the original JLP bound, one would have $\pi_\text{max}\sim\sqrt{\frac{ \mathcal{V}\cdot 2 a E}{\epsilon}}$. We compare $\pi_\text{max}$ obtained from the two methods in Fig. \ref{fig:phi4_m0p5_lamb16_pi_bound_compare}. As shown in the figure, while $\pi_\text{max}$ in our method also grows with the volume, it is nevertheless much smaller than the $\pi_\text{max}$ in original JLP bound at large volumes.

\subsection{Bounding $R$ field in 2+1D U(1) gauge theory in dual formalism\label{sec:results:R}}
\begin{figure}[h!]
    \centering
    \includegraphics[width=0.9\linewidth]{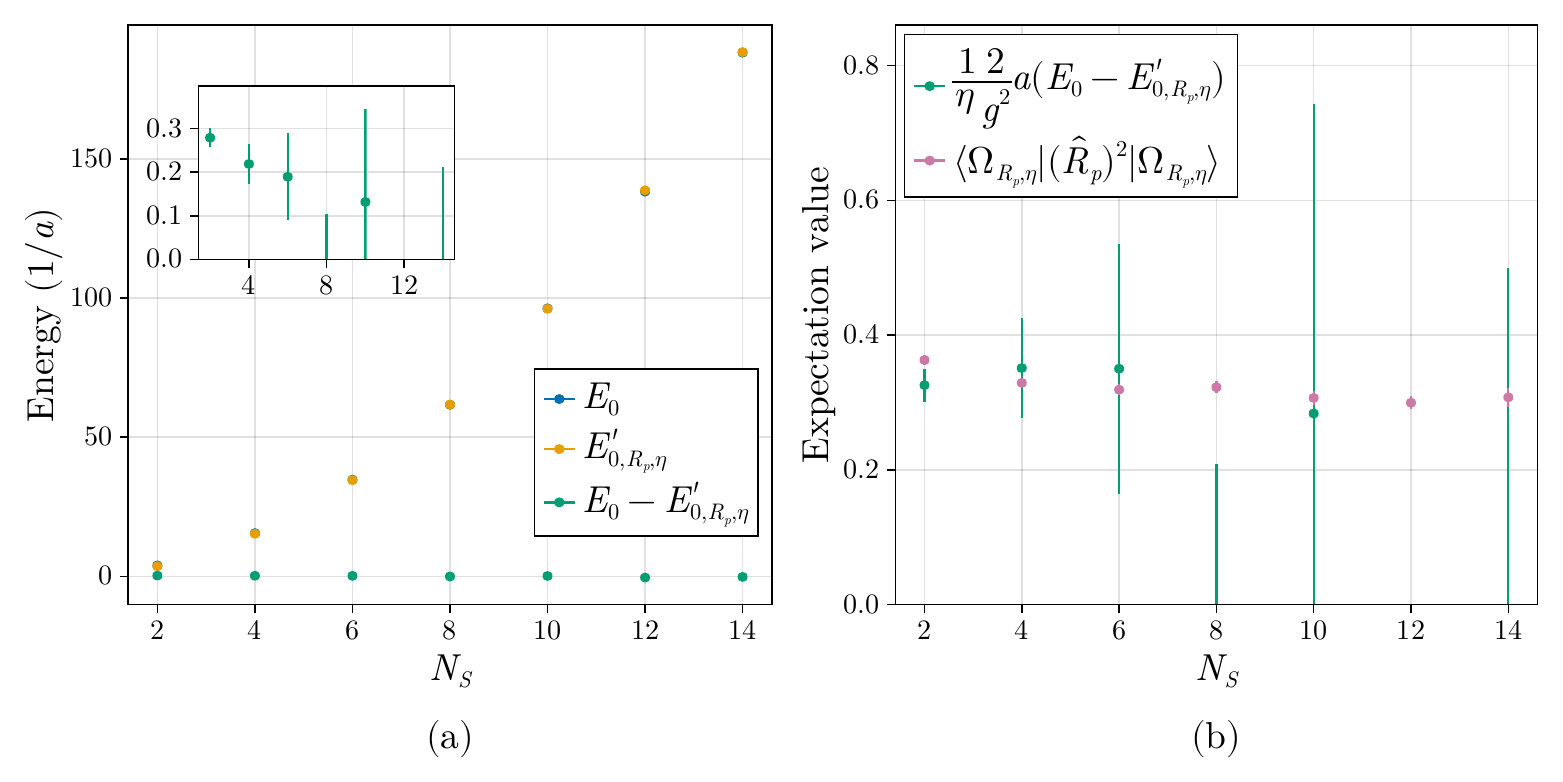}
    \caption{Comparison between energy scales. (a) the ground state energy $E_0$ of $\hat H_{\text{U(1)}}$,  $E'_{0,R_p,\eta}$ of $\hat H'_{R_p,\eta}$, and their difference $E_0-E'_{0,R_p,\eta}$ at different lattice size $N_S$. Each lattice has $N_S\times N_S$ plaquettes spatially.  The inset plot shows a zoomed-in view of $E_0-E'_{0,R_p,\eta}$.
    (b) $\frac{1}{\eta}\frac{2}{g^2}a(E_0-E'_{0,R_p,\eta})$ and its upper bound $\bra{\Omega_{R_p,\eta }} ( \hat R_p )^2 \ket{\Omega_{R_p,\eta }}$. }
    \label{fig:dual_g1_R_corner_energy_plot}
\end{figure}

Using the method in Section \ref{sec:methodII:R}, we can obtain the truncation bound for the rotor electric field $R$. As an example, we estimated the required cutoff for $R_p$ at plaquette position $(1,1)$. The cutoff for other positions can be obtained analogously.  
Here, the factor $\eta$ is chosen to be $\eta=\frac{1}{2}\eta_{\text{max}}$, where $\eta_{\text{max}}$ is the maximal $\eta$ such that $\sum_{r,s} \hat R_{r} M_{r,s} \hat R_{s} -\eta (\hat R_p)^2 $ is positive semi-definite, and $\eta_{\text{max}}$  is volume-dependent (see Appendix \ref{app:dual:eta}).  
Other settings are similar to those in Section~\ref{sec:results:B}.
We plot the ground state energy $E_0$ of the Hamiltonian $\hat{H}_{\text{U(1)}}$, the ground state energy  $E'_{R_p,\eta}$ of $\hat H'_{R_p,\eta}$, and their difference $E_0-E'_{0,R_p,\eta}$ in Fig. \ref{fig:dual_g1_R_corner_energy_plot}(a). Then, in Fig. \ref{fig:dual_g1_R_corner_energy_plot}(b), we plot $\frac{1}{\eta}\frac{2}{g^2}a(E_0-E'_{0,R_p,\eta})$  along with its upper bound $\bra{\Omega_{R_p,\eta }} ( \hat R_p )^2 \ket{\Omega_{R_p,\eta }}$. As expected, $E_0$ and $E'_{0,R_p,\eta}$ become large at large volumes. On the other hand, the volume dependence of $E_0-E'_{0,R_p,\eta}$  is very small. $\bra{\Omega_{R_p,\eta }} ( \hat R_p )^2 \ket{\Omega_{R_p,\eta }}$ is nearly constant in Fig.~\ref{fig:dual_g1_R_corner_energy_plot}(b).

\begin{figure}[h!]
    \centering
    \includegraphics[width=0.9\linewidth]{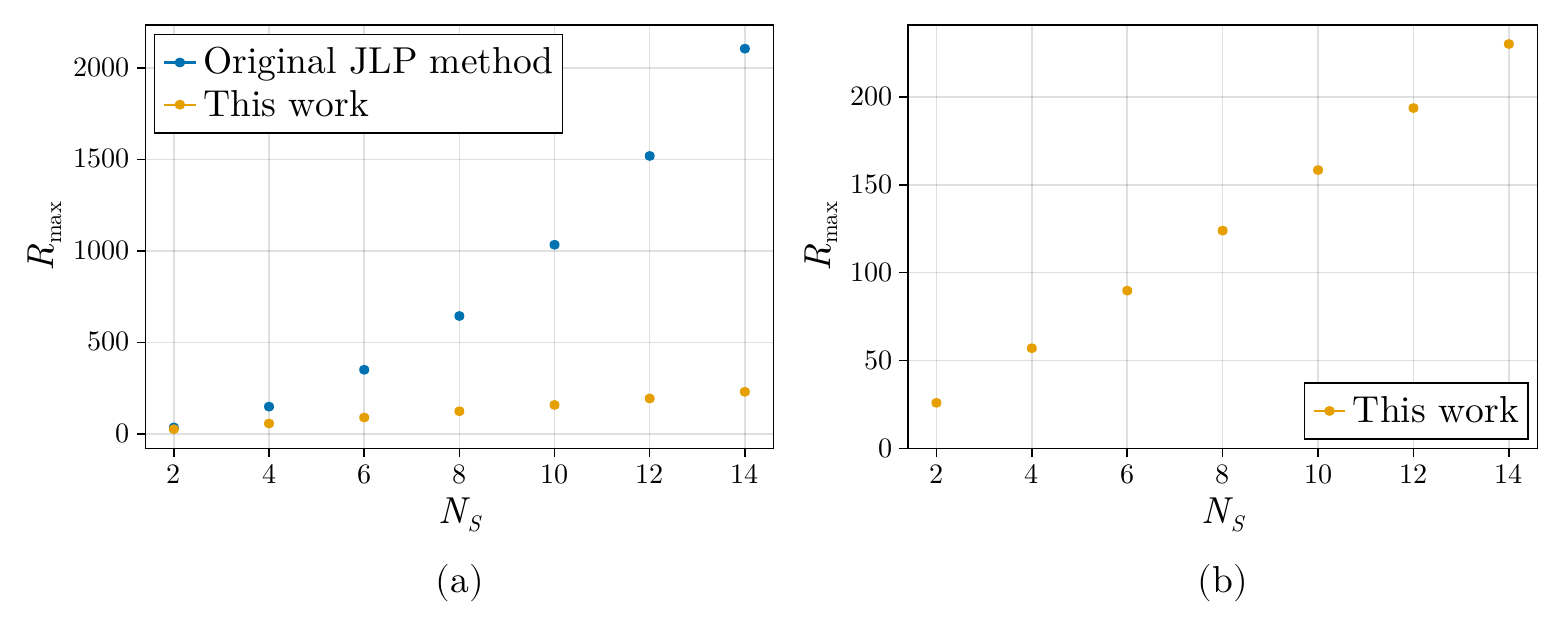}
    \caption{Estimation of the relative scale of truncation $R_p$ for plaquette at position $(1,1)$ at different lattice sizes. Each lattice has $N_S\times N_S$ plaquettes spatially.  Here, $a \Delta E$ is set to $1$.  In (a), we plot the required truncation cutoff $ R_{p,\text{max}}$  as obtained in our method and in the original JLP method. (b) gives a zoomed-in view of our results.}
    \label{fig:dual_g1_R_corner_bound_compare}
\end{figure}

Once $\bra{\Omega_{R_p,\eta }} ( \hat R_p )^2 \ket{\Omega_{R_p,\eta }}$ is evaluated, the cutoff $R_{p,\text{max}}$ can be determined from Eqns. \ref{eqn:Rp_bound_summary} and \ref{eqn:Rp_cutoff}.
In Fig. \ref{fig:dual_g1_R_corner_bound_compare}, we compare the obtained $R_{p,\text{max}}$ of our method with that from the original JLP method. 
As shown in Fig. \ref{fig:dual_g1_R_corner_bound_compare}(b), the bound obtained in our method appears to increase linearly with $N_S$, i.e., scales as $\sqrt{\mathcal{V}}$. On the other hand, as shown in Fig. \ref{fig:dual_g1_R_corner_bound_compare}(a), the bound in the original JLP method displays a considerably faster rate of increase. Hence, the method outlined in Section \ref{sec:methodII:R} can still effectively tighten the cutoff choice for $R$ in this case.

\section{Time evolution cost reduction \label{sec:time_evolution}}

In this section, we discuss the implication of our improved error bounds for Hamiltonian simulation. 
Tightening the field truncation reduces the number of gates required to simulate time evolution in two ways.
First, because fewer qubits are needed per site, the local Hamiltonian terms are simpler and so are the circuits needed for time evolution. 
Note that, because the number of qubits per site $n_q$ scales as $n_q = \mathcal{O}(\log_2 \phi_{\rm max} \pi_{\rm max})$~\cite{jordan_quantum_2012}, combined with the fact that the gate cost scales polynomially with $n_q$, the savings from a reduced qubit count is $\mathcal{O}({\rm polylog}(\mathcal{V}))$.

The second, and more important, reduction stems from the dependence of time-evolution algorithms on the norm of the Hamiltonian terms. 
By justifying a smaller truncation cutoff, the norm of the truncated Hamiltonian is correspondingly reduced by a polynomial factor. 
This reduction tightens the resource estimates for simulation algorithms like Trotterization or quantum signal processing. 
We emphasize that our improved truncation bounds apply to the state truncation error, while the dynamical error introduced by evolving according to the truncated Hamiltonian is left for future work.
Nevertheless, any rigorous reduction of the cutoff must first ensure that the relevant states are faithfully represented, so tightening the state truncation bound is a necessary prerequisite.
Therefore, our improved state truncation cutoff can unlock potential savings of $\mathcal{O}(\mathrm{poly}(\mathcal{V}))$ for time evolution gate cost.

As two representative algorithms, we consider simulation via product formulas (Trotter) and Quantum Signal Processing (QSP) based algorithms.
For simplicity, we consider the asymptotic gate cost reductions from our improved truncation bounds, and leave detailed gate cost analysis for future work; we take the asymptotic gate complexities from, e.g., Refs.~\cite{hariprakash2025strategies, Hardy:2024ric, kane2025block}.

Consider first simulation using product formulas.
The overall simulation cost is determined by multiplying the gate cost of an individual Trotter step by the total number of steps required to achieve time evolution over a duration $t$ with accuracy $\epsilon_{\rm sim}$.
Here, the error is quantified as $\epsilon_{\rm sim} = \|U(t)-S_p(t)\|$, with $U(t)$ denoting the exact time-evolution operator and $S_p(t)$ the corresponding $p$'th order product-formula approximation.
Each Trotter step consists of multiple stages, where each stage typically involves exponentiating the individual terms of the Hamiltonian.
To demonstrate the possible improvements, we consider a first order product-formula.
The required number of Trotter steps $N_\text{steps}$ for a 1st order product formula is bounded by~\cite{childs_theory_2021}
\begin{equation}
    N_\text{steps} \geq \frac{\widetilde{\alpha}_{\rm comm} t^2}{\epsilon_{\rm sim}}\,.
\end{equation}
Here $\widetilde{\alpha}_{\rm comm}$ characterizes the degree of non-commutativity among the Hamiltonian and is given by $\widetilde \alpha_{\rm comm} = \mathcal{O}(\| [\hat H_\phi, \hat H_\pi] \|)$, where $\hat H_{\phi}$ and $\hat H_{\pi}$ contain all the $\hat \phi$ and $\hat \pi$ terms, respectively.
Using the canonical commutation relations, the dominant term in $\widetilde{\alpha}_{\rm comm}$ is 
\begin{equation}
    \widetilde{\alpha}_{\rm comm} = \mathcal{O}\left(\mathcal{V} \left\|[\hat \pi^2, \hat \phi^4]\right\|\right) = \mathcal{O}(\mathcal{V} \pi_{\rm max} \phi_{\rm max}^3)\,.
\end{equation}

As previously mentioned, the number of gates per Trotter step scales as $\widetilde{\mathcal{O}}(\mathcal{V})$.
Using these results, up to multiplicative logarithmic factors, the asymptotic scaling of using 1st order Trotter to simulate time evolution for fixed $t$ and $\epsilon_{\rm sim}$ is 
\begin{equation}
    {\rm Gates}_{\rm PF}(\phi_{\rm max},\pi_{\rm max}, \mathcal{V}) = \widetilde{\mathcal{O}}\left(\mathcal{V}^2 \pi_{\rm max} \phi_{\rm max}^3\right)
\end{equation}

Using the original bounds in Ref.~\cite{jordan_quantum_2012}, $\phi_{\rm max}^{\rm jlp}$ and $\pi_{\rm max}^{\rm jlp}$ scale as $\mathcal{O}(\mathcal{V})$, while using our methods we have (observed) $\phi^{\rm ours}_{\rm max} = \widetilde{\mathcal{O}}(1)$ and $\pi_{\rm max}^{\rm ours}=\mathcal{O}(\sqrt{\mathcal{V}})$. 
The expected asymptotic gate cost reduction in the number of sites from using our improved bounds is therefore
\begin{equation}
    \frac{{\rm Gates}_{\rm PF}(\phi_{\rm max}^{\rm jlp},\pi_{\rm max}^{\rm jlp}, \mathcal{V})}{{\rm Gates}_{\rm PF}(\phi_{\rm max}^{\rm ours},\pi_{\rm max}^{\rm ours}, \mathcal{V})} = \widetilde{\mathcal{O}}\left(\mathcal{V}^{7/2}\right)\,.
\end{equation}
Using similar arguments for the 2+1D U(1) lattice gauge theory, we find asymptotic improvements of
\begin{equation}
    \frac{{\rm Gates}_{\rm PF}(B_{\rm max}^{\rm jlp},R_{\rm max}^{\rm jlp}, \mathcal{V})}{{\rm Gates}_{\rm PF}(B_{\rm max}^{\rm ours},R_{\rm max}^{\rm ours}, \mathcal{V})} = \widetilde{\mathcal{O}}\left(\mathcal{V}^{3/2}\right)\,.
\end{equation}

Another common approach to simulating time-evolution is quantum signal processing (QSP)~\cite{Low:2016sck,Low:2016znh,Gilyen:2018khw,Motlagh:2023oqc}.
The building block in such approaches is a block-encoding of the Hamiltonian $U_H$, where one embeds the Hamiltonian in a larger, unitary, matrix.
To embed $H$ in $U_H$, it must be scaled down by $\beta$ such that $\|H/\beta\| \leq 1$.
This so-called scale factor $\beta$ directly effects the cost of time evolution, and, as we will see, our methods for bounding $\phi_{\rm max}, \pi_{\rm max}$ dramatically reduce the size of the scale factor.

Using QSP, by performing repeated calls to $U_H$ alternated with parameterized single qubit rotations, one approximates $e^{-i H t}$ using a truncated Jacobi-Anger expansion. 
The dominant cost is the queries to $U_H$, which to simulate for time $t$ to error $\epsilon_{\rm sim}$ scales asymptotically as~\cite{Low:2016sck, Low:2016sck}
\begin{equation}
    N_{U_H} = \mathcal{O}\left(\beta t + \log\left( \frac{1}{\epsilon_{\rm sim}}\right)\right)\,.
\end{equation}
If $C_{U_H}$ is the number of gates to construct $U_H$, then the cost of time evolution is $\text{Gates}_{\rm QSP} = N_{U_H} \cdot C_{U_H}$.
Because the scale factor $\beta$ grows with system size, one generally has $\beta t \gg \log \frac{1}{\epsilon_{\rm sim}}$, and the cost is essentially proportional to $\beta t \cdot C_{U_H}$.
This implies that cost reductions for QSP time evolution can be determined by taking ratios of the scale factor $\beta$ multiplied by the number of gates to implement the block-encoding.

From Refs.~\cite{hariprakash2025strategies, Hardy:2024ric, kane2025block}, one can block-encode the scalar field theory Hamiltonian with scale factor whose asymptotic scaling is
\begin{equation}
    \beta = \mathcal{O} \left(\mathcal{V}\left[\pi_{\rm max}^2 + \phi^4_{\rm max} \right]\right)\,.
\end{equation}
This expression shows that reductions in $\pi_{\rm max}$ and $\phi_{\rm max}$ lead directly to reduced simulation cost by reducing $\beta$.

Similar to the Trotter case, the number of gates required to block-encode $H$ scales as $C_{U_H} = \widetilde{\mathcal{O}}(\mathcal{V})$.
The expected asymptotic gate cost reduction can be found by taking ratios of $\beta$, which gives
\begin{equation}
    \frac{{\rm Gates}_{\rm QSP}(\phi_{\rm max}^{\rm jlp},\pi_{\rm max}^{\rm jlp}, \mathcal{V})}{{\rm Gates}_{\rm QSP}(\phi_{\rm max}^{\rm ours},\pi_{\rm max}^{\rm ours}, \mathcal{V})} = \widetilde{\mathcal{O}}\left(\mathcal{V}^3\right)\,.
\end{equation}
In a similar way, the asymptotic improvements for the 2+1D U(1) lattice gauge theory are
\begin{equation}
    \frac{{\rm Gates}_{\rm QSP}(B_{\rm max}^{\rm jlp},R_{\rm max}^{\rm jlp}, \mathcal{V})}{{\rm Gates}_{\rm QSP}(B_{\rm max}^{\rm ours},R_{\rm max}^{\rm ours}, \mathcal{V})} = \widetilde{\mathcal{O}}\left(\mathcal{V}\right)\,.
\end{equation}

To summarize, while our improved truncation strategies only result in ${\rm polylog}(\mathcal{V})$ reductions in the number of qubits per bosonic mode, our improved bounds lead to ${\rm poly}(\mathcal{V})$ asymptotic reductions in the number of gates required to simulate time evolution of the truncated Hamiltonian using both Trotter methods and QSP.
Because the cost of a given time evolution algorithm generally depends on the norm of the Hamiltonian terms, we expect similar gate reductions for time evolution broadly.

\section{Conclusion \label{sec:conclude}}

In this work, we analyze the energy-based boson truncation error bound introduced by Jordan, Lee, and Preskill (JLP)~\cite{jordan_quantum_2012} and develop a strategy to tighten it. We identify two sources of looseness in the original JLP bound. First, many terms are unnecessarily discarded in the original derivation of the energy-based bound. 
Second, the lack of detailed information on the field-value distribution leads to the use of pessimistic bounds.
Incidentally, we also note an often underappreciated feature of the original result: beyond the explicit $\sqrt{\mathcal{V}}$ factor in the truncation cutoff $\phi_{\text{max}}\sim \sqrt{\frac{\mathcal{V} E}{m_0^2 \epsilon}}$, there is typically an additional implicit $\sqrt{\mathcal{V}}$ contribution arising from $\sqrt{E}$, which makes the volume scaling of $\phi_{\text{max}}$ worse than it may initially appear. 

To tighten the bound, we introduce two complementary techniques: a numerical ``Monte Carlo trick'' and an analytical ``$p$-norm trick.'' The Monte Carlo trick tightens the inequality by numerically bounding the discarded terms in the original derivation of the JLP bound, while the $p$-norm trick exploits global information about the field's probability distribution rather than treating each lattice site independently. These two methods can often be combined, as in the case of the $\phi$ field, leading to substantial improvements at large volumes. At the same time, each method can also be useful on its own. In particular, in certain cases, the $p$-norm trick can serve as a standalone technique to gain a ``free'' factor of $\sqrt{\mathcal{V}}$ improvement over the original JLP bound with minimal additional effort.

Applying this framework substantially reduces the required boson truncation cutoffs. For the $\phi$ field in the 1+1D $\phi^4$ theory, our numerical results demonstrate that, for a fixed error budget $\epsilon$, the required cutoff $\phi_{\text{max}}$ exhibits a very mild dependence on $\mathcal{V}$, appearing to scale logarithmically or algebraically with a low exponent, rather than linearly. Relative to the original energy-based bound in Ref.~\cite{jordan_quantum_2012}, this corresponds to nearly a factor of $\mathcal{V}$ improvement. For the conjugate field $\pi$, however, the currently available techniques yield more modest gains. In this case, the resulting cutoff $\pi_\text{max}$ scales as $\sqrt{\mathcal{V}}$. Compared to the linear $\mathcal{V}$ dependence of the original JLP bound, this still represents a factor $\sqrt{\mathcal{V}}$ reduction. Furthermore, our methodology is not limited to the $\phi^4$ scalar theory and can be applied to other theories. As an application to gauge theory, we apply our method to the 2+1D non-compact U(1) gauge theory in the dual formalism, obtaining improved truncation cutoffs for the magnetic field $B$ and electric rotor field $R$. The resulting reductions are similar to those obtained for the $\phi$ and $\pi$ fields in the scalar field theory, respectively.

Although our approach makes use of classical Monte Carlo calculations, it is not inherently “more numerical” than the original JLP bound.
Notably, the original bound also requires numerical input: the energy $E$ appearing in the bound includes the vacuum energy $E_0$, which generally must be evaluated numerically in the absence of an exact solution.
From this perspective, our work does not introduce additional numerical requirements; rather, it clarifies the necessity of such calculations.

Because our approach builds on the energy-based framework of Ref.~\cite{jordan_quantum_2012}, it inherits some of its limitations. First, the error bound decays only polynomially with respect to the truncation cutoff, which is not asymptotically optimal when compared to alternative methods. Second, the analysis focuses on the fidelity of the truncated quantum state and does not account for the dynamical error introduced by truncating the Hamiltonian. 
Extending the present results beyond the scope of the current analysis remains an important direction for future work.

Additionally, while the present work focuses on purely bosonic theories, many lattice field theories of interest involve fermionic degrees of freedom. A natural next step is therefore to extend the improved energy-based bound to theories containing fermions. In principle, the core ideas introduced in this work can be carried over to such settings, although this requires handling the additional complications that arise from a fermionic path integral. We plan to pursue these extensions in future work.

Finally, we emphasize that the energy-based bound is not mutually exclusive with other truncation methods but can instead complement them. For example, Ref.~\cite{tong_provably_2022} derives bounds on quantum number growth during time evolution, assuming a bounded initial state; energy-based methods can help establish such bounds for the initial state itself. More broadly, we hope that the techniques introduced in this work will inspire new ideas and prove useful in related studies.

\begin{acknowledgments}
We would like to thank Zohreh Davoudi for helpful discussions and Norbert Linke for his perceptive comments. 
JY is supported by the National Quantum Laboratory (QLab) at the University of Maryland.
The work of C.F.K. was supported by the U.S. Department of Energy (DOE), Office of Science, Accelerated Research in Quantum Computing, Fundamental Algorithmic Research toward Quantum Utility (FAR-Qu); the U.S. DOE, Office of Science, Office of Nuclear Physics (award no. DE-SC0026067); and in part by the Department of Physics, Maryland Center for Fundamental Physics, and the College of Computer, Mathematical, and Natural Sciences at the University of Maryland, College Park.
\end{acknowledgments}

\section*{Data availability}
The source code used in our study can be found in Ref.~\cite{boson_truncation_source_code}.

\FloatBarrier
\appendix

\section{2+1D non-compact U(1) gauge theory in dual formalism \label{app:dual}}
\begin{figure}[h!]
    \centering
    \begin{tikzpicture}[scale=2.5, >=Stealth]

    \draw[thick, gray] (0,0) grid (2,2);

    \draw[thick, black, ->] (0,0) -- (0.95, 0) node[midway, below=3pt] {$E^x_{(0,0)}$};
    \draw[thick, black, ->] (0,0) -- (0, 0.95) node[midway, left=3pt] {$E^y_{(0,0)}$};
    \draw[thick, black, ->] (1,0) -- (1.95, 0) node[midway, below=3pt] {$E^x_{(1,0)}$};

    \foreach \x/\y/\name in {0.5/0.5/R_1, 1.5/0.5/R_2, 0.5/1.5/R_3, 1.5/1.5/R_4} {
        \node at (\x, \y) {$\name$};
        \draw[thick, blue, ->] (\x+0.25, \y) arc (0:270:0.25);
    }

    \foreach \x in {0, 1, 2} {
        \foreach \y in {0, 1, 2} {
            \filldraw[black] (\x,\y) circle (1.5pt);
        }
    }

    \node[below left] at (0,0) {(0,0)};
    \node[below]      at (1,0) {(1,0)};
    \node[below right] at (2,0) {(2,0)};
    \node[left]       at (0,1) {(0,1)};
    \node[left]       at (0,2) {(0,2)};

\end{tikzpicture}
    \caption{An illustration of a lattice system of $2\times 2$ plaquettes with open boundary condition.}
    \label{fig:dual_plaquettes}
\end{figure}
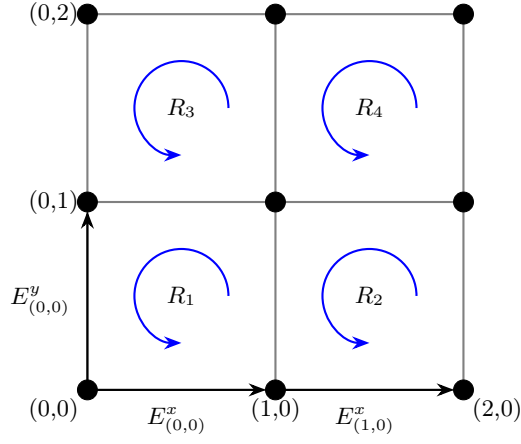

In this section, we include a brief introduction to the 2+1D  U(1) gauge theory and describe how the dual formalism~\cite{bender_gauge_2020,bauer_efficient_2021,kane_efficient_2022} can relate to the Kogut-Susskind formalism~\cite{kogut_hamiltonian_1975}.
In Kogut-Susskind formalism, the degrees of freedom for the gauge field reside on the gauge links. On each link, the gauge field is described by the gauge link $U_l$ and the electric field $E_l$, with commutation relation $[\hat E_l, \hat U_m]=\delta_{l,m} \hat U_m$. 
By contrast, the dual formalism uses the plaquette degrees of freedom $B_p$ and $R_p$. They can be related to $U_l$ and $E_l$ as follows. One can define a plaquette Wilson loop $\hat P_{\mathbf{n}}$ as $\hat P_{\mathbf{n}}=\hat U_{\mathbf{n}}^x \hat U_{\mathbf{n}+\hat{e}_x}^y (\hat U_{\mathbf{n}+\hat{e}_j}^x)^\dagger (\hat U_{\mathbf{n}}^y)^\dagger$. Then, for a plaquette $p$, one can define $\hat B_p$ as $\hat{P}_p = e^{i \hat B_p}$. 
At the same time, $R$ is known as the rotator electric field, and can be expressed as linear combinations of the electric field on gauge links. In the example shown in Fig. \ref{fig:dual_plaquettes}, one has $\hat E^x_{(0,0)} = \hat R_1$, $\hat E^y_{(0,0)} =-\hat R_1$, $\hat E^y_{(1,0)}=\hat R_1-\hat R_2$, $\hat E^x_{(0,1)}=-\hat R_1+\hat R_3$, etc. 
With such definitions, one can see $R$ and $B$ follow commutation relationship $[\hat B_p,\hat R_q]= i \delta_{p,q}$.
One can also write down the Hamiltonian for the dual formalism based on the Kogut-Susskind Hamiltonian, which has two gauge field terms $\hat H_B$ and $\hat H_E$, with $\hat H_B = \frac{1}{a}\frac{1}{2 g^2} \sum_p \Tr[2 - \hat P_p - \hat P_p^\dagger]$ and $\hat{H}_E = \frac{1}{a}\frac{g^2}{2} \sum_l \hat E_l^2$. For the magnetic field term, $\hat H_B = \frac{1}{a}\frac{1}{2 g^2} \sum_p \Tr[2 - \hat P_p - \hat P_p^\dagger]$, one can rewrite it as $\hat H_B =\frac{1}{a}\frac{1}{ g^2}\sum_p \left(1-\cos(\hat B_p)\right)  $, or alternatively $\hat H_B = \frac{1}{2 g^2} \sum_p \hat B_p^2$. The former is refered to as the compact formulation, while the latter is refered to as the non-compact formulation~\cite{bauer_efficient_2021}. Both should agree in the continuum limit.
The electric Hamiltonian $\hat{H}_E = \frac{1}{a}\frac{g^2}{2} \sum_l \hat E_l^2$ can likewise be expressed  in terms of $R_p$. In the example given in Fig. \ref{fig:dual_plaquettes}, the electric Hamiltonian can be written as 
\begin{equation}
    \hat H_E = \frac{1}{a}\frac{g^2}{2} \left( 4(\hat R_1^2+\hat R_2^2+\hat R_3^2+\hat R_4^2)-2(\hat R_1 \hat R_2 + \hat R_1 \hat R_3+\hat R_2 \hat R_4 +\hat R_3 \hat R_4)\right).
\end{equation}
In a more concise fashion, one may write $\hat H = \frac{1}{a}\frac{g^2}{2} \sum_{p,q} \hat R_p M_{p,q} \hat R_q$, where the matrix $M_{p,q}$ in this example is
$$
\begin{pmatrix}
    4   & -1 & -1 & 0 \\ 
    -1 &  4  &   0 & -1 \\
    -1 &  0  &   4 & -1 \\
    0   & -1 & -1 & 4
\end{pmatrix}.
$$
In more general cases, $M_{p,q}$ can also be obtained straight-forwardly from the relationship between $E$ and $R$ field. It is not hard to see that $M_{p,q}$ will always be a real symmetric positive-definite matrix.

To summarize, in the non-compact version of the dual formalism, the pure gauge Hamiltonian can be written as
\begin{equation}
\hat H_{\text{U(1)}} = \frac{1}{a} \frac{g^2}{2} \sum_{p,q} \hat R_{p} M_{p,q} \hat R_{q}+\frac{1}{a}\frac{1}{2 g^2} \sum_{p} \left(\hat B_{p}\right)^2 .
\end{equation}

For context, we note one motivation discussed in previous work on the dual formulation~\cite{bauer_efficient_2021}. In the weak-coupling regime, the wave function width in the magnetic basis is argued to scale as $\sim g$, suggesting that the magnetic field distribution is localized near $B=0$ and might permit a relatively small truncation cutoff. This differs from the standard Kogut-Susskind formulation, which uses compact gauge-link variables defined modulo $2\pi$ across all couplings. In the present work, we do not attempt a formulation-by-formulation resource comparison, but instead use the non-compact dual formulation as a gauge-theory example to demonstrate different aspects of the improved energy-based truncation bound.

\subsection{Property of $M_{p,q}$ matrix at large lattice size\label{app:dual:eta}}
To be able to use the energy-based bound for the dual formalism U(1) gauge theory, the properties of the $M_{p,q}$ matrix needs to be studied. 

\begin{figure}[h!]
    \centering
    \includegraphics[width=0.5\linewidth]{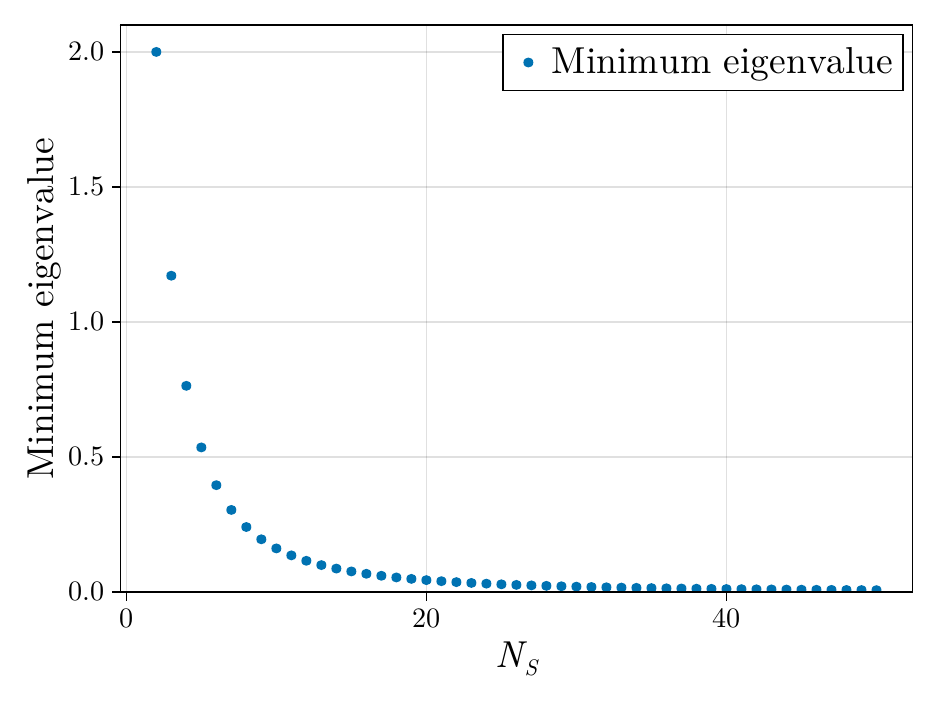}
    \caption{Minimum eigenvalue for $M$ matrix for different lattice sizes in 2+1d dual formalism U(1) gauge theory.}
    \label{fig:dual_M_mat_lambda.pdf}
\end{figure}

As mentioned in Section \ref{sec:methodII:R}, the maximal $\lambda_{\text{max}}$ such that $\hat{H}_{\text{U(1)}}-\lambda_{\text{max}}\frac{1}{a}\frac{g^2}{2} \sum_p (\hat R_p)^2$ is positive-semidefinite will decrease considerably fast as the lattice size increases. Here, we will discuss the reasons. To make sure the modified Hamiltonian is positive-definite, the electric part needs to be lower bounded. In other words, $M- \lambda_{\text{max}} I$ should be lower bounded. Correspondingly, $\lambda_{\text{max}}$ is the minimum eigenvalue of the $M$ matrix.  In Fig. \ref{fig:dual_M_mat_lambda.pdf}, the minimum eigenvalue is plotted against the lattice size $N_S$, where the lattice has $N_S\times N_S$ plaquettes. As shown in the figure, the minimum eigenvalue of $M$ decreases considerably fast as $N_S$ increases; hence, the bound $\bra{\psi} (\hat R^{(2)})^2\ket{\psi}\leq \frac{1}{\lambda_{\text{max}}} \frac{2}{g^2} a E$ is unsuitable to use for large lattice size.

\begin{figure}[h!]
    \centering
    \includegraphics[width=0.5\linewidth]{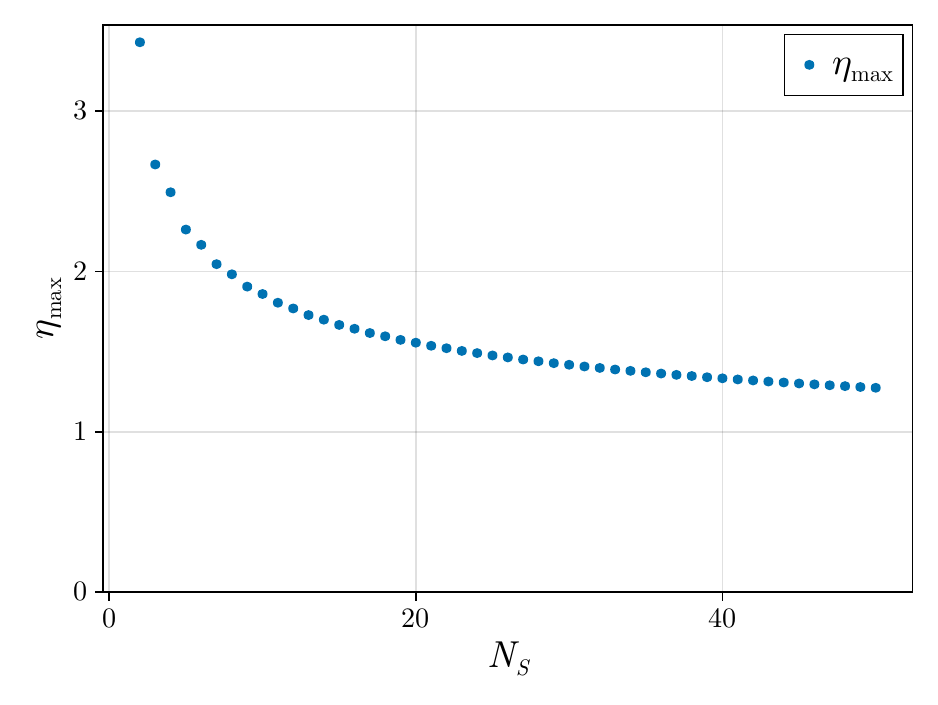}
    \caption{$\eta_{\text{max}}$ for different lattice sizes in 2+1d dual formalism U(1) gauge theory.}
    \label{fig:dual_M_mat_etamax}
\end{figure}

On the other hand, to use the proposed methodology as described in Section \ref{sec:methodII:R}, one needs to ensure that $M_{r,s}-\eta \delta_{r,p} \delta_{p,s}$ is positive-semidefinite for a chosen plaquette $p$. To derive a bound for maximum $\eta$, note that we can write
\begin{equation}
    M - \eta e_p e_p^T = M^{1/2} \left( I - \eta M^{-1/2} e_p e_p^T M^{-1/2}\right)M^{1/2}, 
\end{equation}
and that the rank-1 matrix $ M^{-1/2} e_p e_p^T M^{-1/2}$ has a single eigenvalue being $e_p^T M^{-1} e_p$, i.e., $(M^{-1})_{p,p}$. Thus, to make sure $M - \eta e_p e_p^T $ has only non-negative eigenvalues, one needs to make sure $\eta (M^{-1})_{p,p} \leq 1 $. To make sure the inequality holds for all plaquettes, one can choose
\begin{equation}
    \eta_{\text{max}} = 1 / \max(\text{diag}(M^{-1})),
\end{equation}
where $\text{diag}$ denotes taking the diagonal entries of the matrix. In Fig.~\ref{fig:dual_M_mat_etamax}, $\eta_{\text{max}}$ is plotted against different $N_S$. As shown in the figure, $\eta_{\text{max}}$ decays slowly at large $N_S$. Hence, the bound given by Eqn. \ref{eqn:Rp_bound_summary} will still be useful at large $N_S$.

\section{Derivation of path integral\label{app:path_integral}}

In this section, we describe how one can obtain the path integral expression used in the main text, e.g., Eqns.~\ref{eqn:scalar_S_unmodified}, \ref{eqn:dual_R_action}, and \ref{eqn:dual_R_R2tilde}, as well as how one can calculate the vacuum energy. We will summarize the main ideas here. For additional reference, a pedagogical discussion of the path integral formulation can be found in standard lattice field theory texts~\cite{gattringer_quantum_2010,creutz_quarks_1983,montvay_quantum_1997}. 

Consider a Hamiltonian $\hat{H}$ and denote its eigenstates by $\ket{n}$, with $\ket{0}$ being its ground state. Given an operator $\hat{O}$ and a large positive time scale $T$, we have
\begin{equation}
\begin{aligned}
    \frac{\Tr[e^{-\hat H T/2} \hat O e^{-\hat H T/2}]}{\Tr[e^{-\hat H T}]} =& \frac{\sum_n\bra{n}e^{-\hat H T/2} \hat O e^{-\hat H T/2} \ket{n}}{\sum_n \bra{n}e^{-\hat H T} \ket{n}}\\
    =& \frac{\sum_n e^{-E_n T}\bra{n}\hat O \ket{n}}{\sum_n e^{-E_n T}} \\
    =& \frac{\bra{0}\hat O \ket{0}+\sum_{n\neq 0} e^{-(E_n-E_0) T}\bra{n}\hat O \ket{n}}{1+\sum_{n\neq 0} e^{-(E_n-E_0) T}} \\
    \approx& \bra{0}\hat O \ket{0} + \mathcal{O}\left(e^{-E_g T}\right),
\end{aligned}
\end{equation}
where $E_g$ is the energy difference between the ground state and the excited states, and exact equality can be obtained in the limit $T\rightarrow \infty$. In this way, one can relate the ground state expectation value of $\hat O$ to the object $\Tr[\hat{O} e^{-\hat H T}]$ and $\Tr[e^{-\hat H T}]$, which have a path-integral interpretation. In the following two subsections, we will review how $\Tr[\hat{O} e^{-\hat H T}]$ and $\Tr[e^{-\hat H T}]$ translate to path integral in the case of the scalar field theory and the 2+1D U(1) gauge theory in dual formalism.

\subsection{Path integral in scalar field theory \label{app:path_integral:scalar}}
Here, let's first consider $\Tr[e^{-\hat H T}]$. To convert this into the form of a path integral, we can insert complete bases of states at different time slices. Firstly, suppose $T=N a_0$, where $a_0$ can be interpreted as the temporal lattice spacing, then one can write $e^{-\hat H T}=(e^{-\hat H a_0})^N$. Then, notice that the identity operator can be written as
\begin{equation}
    I=\int \mathcal{D} \boldsymbol{\phi} \ket{\boldsymbol{\phi}}\bra{\boldsymbol{\phi}},
    \label{eqn:identity-phi}
\end{equation}
where $\mathcal{D} \boldsymbol{\phi}=\prod_\mathbf{x} \dd \phi(\mathbf{x})$, and $\ket{\boldsymbol{\phi}}$ is a shorthand for
\begin{equation}
    \ket{\boldsymbol{\phi}} = \ket{\phi(\mathbf{x}_1)}\otimes \ket{\phi(\mathbf{x}_2)} \otimes ...\otimes \ket{\phi(\mathbf{x}_\mathcal{V})}.
\end{equation}
Hence, one can write
\begin{equation}
    \begin{aligned}
        \Tr[e^{-\hat H T}] = & 
        \Tr\left[\left(\int \mathcal{D} \boldsymbol{\phi} \ket{\boldsymbol{\phi}}\bra{\boldsymbol{\phi}} e^{-\hat H a_0}\right)^N\right] \\
        =& \int \mathcal{D} \boldsymbol{\phi}_1\mathcal{D} \boldsymbol{\phi}_2 ...\mathcal{D} \boldsymbol{\phi}_{N} \bra{\boldsymbol{\phi}_N}e^{-\hat H a_0} \ket{\boldsymbol{\phi}_{N-1}}\bra{\boldsymbol{\phi}_{N-1}}e^{-\hat H a_0} \ket{\boldsymbol{\phi}_{N-2}} ...\bra{\boldsymbol{\phi}_{2}}e^{-\hat H a_0} \ket{\boldsymbol{\phi}_{1}}\bra{\boldsymbol{\phi}_{1}}e^{-\hat H a_0} \ket{\boldsymbol{\phi}_{N}}.
    \end{aligned}
    \label{eqn:pi_divide_into_slices}
\end{equation}
Here, the subscript $j$ for $\boldsymbol{\phi}_j$ denotes the temporal slice, not  the spatial index. The next step is to evaluate matrix elements of the form $\bra{\boldsymbol{\phi}_{j+1}}e^{-\hat H a_0} \ket{\boldsymbol{\phi}_{j}}$. To do this, first notice that at small $a_0$, one can write 
\begin{equation}
    e^{-\hat H a_0} \approx e^{-\hat H_{\text{kin}} a_0}e^{-\hat H_{\text{pol}} a_0}+\mathcal{O}(a_0^2),
\end{equation}
where
\begin{equation}
    \begin{aligned}
        \hat H_{\text{kin}} = & \frac{1}{a}\sum_{\mathbf{x}}  \frac{1}{2} \hat \pi(\mathbf{x})^2 \\
\hat H_{\text{pol}}        =& \frac{1}{a}\sum_{\mathbf{x}}  \left(\frac{1}{2}(\mathbf{\nabla} \hat\phi)^2(\mathbf{x})+\frac{1}{2} m_0^2 \hat\phi(\mathbf{x})^2+\frac{\lambda_0}{4!}\hat\phi(\mathbf{x})^4\right).
    \end{aligned}
\end{equation}
Then, note that a given state can alternatively be expanded in the $\pi$ field basis, and similar to Eqn. \ref{eqn:identity-phi}, one can write $I=\int \mathcal{D} \boldsymbol{\pi} \ket{\boldsymbol{\pi}}\bra{\boldsymbol{\pi}}$. From the commutation relationship, one can find $\phi$ and $\pi$ to be conjugate variables; thus, one can write
\begin{equation}
    \bra{\boldsymbol{\phi}}\ket{\boldsymbol{\pi}} = \prod_{\mathbf{x}}\frac{1}{\sqrt{2\pi}} e^{i \pi(\mathbf{x}) \phi(\mathbf{x})}.
\end{equation}
Thus, we have
\begin{equation}
    \begin{aligned}
      &   \bra{\boldsymbol{\phi}_{j+1}}e^{-\hat H a_0} \ket{\boldsymbol{\phi}_{j}} \\
        \approx&  \int \mathcal{D} \boldsymbol{\pi} \bra{\boldsymbol{\phi}_j}\ket{\boldsymbol{\pi}}\bra{\boldsymbol{\pi}}e^{-\hat H_{\text{kin}} a_0}e^{-\hat H_{\text{pol}} a_0}\ket{\boldsymbol{\phi}_{j-1}} \\
        =&  \int \mathcal{D} \boldsymbol{\pi}   \bra{\boldsymbol{\phi}_{j+1}}\ket{\boldsymbol{\pi}}
        \bra{\boldsymbol{\pi}}\ket{\boldsymbol{\phi}_{j}}
        e^{-\frac{a_0}{a}\sum_{\mathbf{x}}  \frac{1}{2} \pi(\mathbf{x})^2}
        e^{-\frac{a_0}{a}\sum_{\mathbf{x}}  \left(\frac{1}{2}(\mathbf{\nabla} \phi_{j})^2(\mathbf{x})+\frac{1}{2} m_0^2 \phi_{j}(\mathbf{x})^2+\frac{\lambda_0}{4!}\phi_{j}(\mathbf{x})^4\right)} \\
        =& \int \mathcal{D} \boldsymbol{\pi}  \left(\frac{1}{2\pi}\right)^{\mathcal{V}}
        e^{-\frac{a_0}{a}\sum_{\mathbf{x}}  \frac{1}{2} \pi(\mathbf{x})^2+i \sum_{\mathbf{x}} \pi(\mathbf{x}) \left(\phi_{j+1}(\mathbf{x})-\phi_{j}(\mathbf{x})\right)}
        e^{-\frac{a_0}{a}\sum_{\mathbf{x}}  \left(\frac{1}{2}(\mathbf{\nabla} \phi_{j})^2(\mathbf{x})+\frac{1}{2} m_0^2 \phi_{j}(\mathbf{x})^2+\frac{\lambda_0}{4!}\phi_{j}(\mathbf{x})^4\right)} \\
        =& \left(\sqrt{\frac{a}{2\pi a_0}}\right)^{\mathcal{V}}
        e^{-\frac{a}{a_0} \sum_{\mathbf{x}}\frac{1}{2}\left(\phi_{j+1}(\mathbf{x})-\phi_{j}(\mathbf{x})\right)^2-\frac{a_0}{a}\sum_{\mathbf{x}}  \left(\frac{1}{2}(\mathbf{\nabla} \phi_{j})^2(\mathbf{x})+\frac{1}{2} m_0^2 \phi_{j}(\mathbf{x})^2+\frac{\lambda_0}{4!}\phi_{j}(\mathbf{x})^4\right)}.
    \end{aligned}
    \label{eqn:pi_single_slice}
\end{equation}
This equation is the basis of the path integral representation. To make things more explicit, we will recognize $a_0$ as the Euclidean time interval, and instead of $\phi_j(\mathbf{x})$, we will write $\phi(t,\mathbf{x})$, where $t$ is the Euclidean time at slice $j$. Putting Eqn. \ref{eqn:pi_single_slice} into Eqn. \ref{eqn:pi_divide_into_slices}, one may obtain
\begin{equation}
    \Tr[e^{-\hat H T}] \approx \mathcal{N} \int \mathcal{D} \phi  \;
    e^{-\sum_{t,\mathbf{x}}\left[\frac{a}{a_0} \frac{1}{2}\left(\phi(t+a_0,\mathbf{x})-\phi(t,\mathbf{x})\right)^2+\frac{a_0}{a}\left(\frac{1}{2}(\mathbf{\nabla} \phi)^2(t,\mathbf{x})+\frac{1}{2} m_0^2 \phi(t,\mathbf{x})^2+\frac{\lambda_0}{4!}\phi(t,\mathbf{x})^4\right)\right]},
    \label{eqn:pi_scalar_S_derive}
\end{equation}
where $\mathcal{N}$ is a normalization factor.
Thus, the action of the theory can be defined as
\begin{equation}
 S[\phi] = \sum_{t, \mathbf{x}} \left[ \frac{a}{a_0}
\frac{(\phi (t + a_0, \mathbf{x}) - \phi (t,
\mathbf{x}))^2}{2} + \frac{a_0}{a} \frac{1}{2}
(\mathbf{\nabla} \phi)^2 (t, \mathbf{x}) +
\frac{a_0}{a} \frac{m_0^2 \phi (t, \mathbf{x})^2}{2} +
\frac{a_0}{a} \frac{\lambda_0}{4!} \phi (t, \mathbf{x})^4 
\right],
\label{eqn:path_int_scalar_S_unmodified}
\end{equation}
as in Eqn. \ref{eqn:scalar_S_unmodified}, and $ \Tr[e^{-\hat H T}]$ can be expressed as $\Tr[e^{-\hat H T}] \approx\mathcal{N} \int \mathcal{D} \phi  \;    e^{- S[\phi]}$.

In a similar fashion, the modified actions in Eqns. \ref{eqn:scalar_S_phi_x_i} and \ref{eqn:scalar_S_phimax} can be obtained, since only the potential term differs.

For $\Tr[\hat{O} e^{-\hat H T}]$, if $\hat{O}$ can be expressed entirely in terms of the $\hat \phi$ field, then its evaluation is straight-forward. Notice
\begin{equation}
    \bra{\boldsymbol{\phi}_{j+1}}\hat O [\hat\phi]e^{-\hat H a_0} \ket{\boldsymbol{\phi}_{j}} = O[\boldsymbol{\phi}_{j+1}] \bra{\boldsymbol{\phi}_{j+1}}e^{-\hat H a_0} \ket{\boldsymbol{\phi}_{j}}.
\end{equation}
Then, using Eqn. \ref{eqn:pi_divide_into_slices}, one can show that
\begin{equation}
    \Tr[\hat O [\hat\phi] e^{-\hat H T}] \approx \mathcal{N} \int \mathcal{D} \phi \;  O[\boldsymbol{\phi}(t_0)] 
    e^{-\sum_{t,\mathbf{x}}\left[\frac{a}{a_0} \frac{1}{2}\left(\phi(t+a_0,\mathbf{x})-\phi(t,\mathbf{x})\right)^2+\frac{a_0}{a}\left(\frac{1}{2}(\mathbf{\nabla} \phi)^2(t,\mathbf{x})+\frac{1}{2} m_0^2 \phi(t,\mathbf{x})^2+\frac{\lambda_0}{4!}\phi(t,\mathbf{x})^4\right)\right]},
    \label{eqn:path_int_scalar_O_phi}
\end{equation}
where $\boldsymbol{\phi}(t_0)$ denotes the $\phi$ field values at a time slice for some time $t_0$. The equality becomes exact in the limit $a_0\rightarrow 0$.

For the $\hat\pi$ field, the calculation is less straight-forward. However, the expectation value can still be evaluated when $\hat O$ takes a simple form, e.g., $\hat \pi(\mathbf{x}_k)^2$, where $\mathbf{x}_k$ is a spatial site. To evaluate it, notice that
\begin{equation}
    \begin{aligned}
         &  \bra{\boldsymbol{\phi}_{j+1}}\hat \pi(\mathbf{x}_k)^2 e^{-\hat H a_0} \ket{\boldsymbol{\phi}_{j}} \\
         \approx & \int \mathcal{D} \boldsymbol{\pi} \bra{\boldsymbol{\phi}_j}\ket{\boldsymbol{\pi}}\bra{\boldsymbol{\pi}}\hat \pi(\mathbf{x}_k)^2 e^{-\hat H_{\text{kin}} a_0}e^{-\hat H_{\text{pol}} a_0}\ket{\boldsymbol{\phi}_{j-1}} \\
         =&  \int \mathcal{D} \boldsymbol{\pi}  \left(\frac{1}{2\pi}\right)^{\mathcal{V}} \pi(\mathbf{x}_k)^2
        e^{-\frac{a_0}{a}\sum_{\mathbf{x}}  \frac{1}{2} \pi(\mathbf{x})^2+i \sum_{\mathbf{x}} \pi(\mathbf{x}) \left(\phi_{j+1}(\mathbf{x})-\phi_{j}(\mathbf{x})\right)}
        e^{-\frac{a_0}{a}\sum_{\mathbf{x}}  \left(\frac{1}{2}(\mathbf{\nabla} \phi_{j})^2(\mathbf{x})+\frac{1}{2} m_0^2 \phi_{j}(\mathbf{x})^2+\frac{\lambda_0}{4!}\phi_{j}(\mathbf{x})^4\right)} \\
        =&\int \mathcal{D} \boldsymbol{\pi}  \left(\frac{1}{2\pi}\right)^{\mathcal{V}} \left(\pi(\mathbf{x}_k)+i \frac{a}{a_0} \left(\phi_{j+1}(\mathbf{x}_k)-\phi_{j}(\mathbf{x}_k)\right)\right)^2
        e^{-\frac{a_0}{a}\sum_{\mathbf{x}}  \frac{1}{2} \pi(\mathbf{x})^2-\frac{a}{a_0} \sum_{\mathbf{x}}\frac{1}{2}\left(\phi_{j+1}(\mathbf{x})-\phi_{j}(\mathbf{x})\right)^2}\\
    & \quad\quad \times
        e^{-\frac{a_0}{a}\sum_{\mathbf{x}}  \left(\frac{1}{2}(\mathbf{\nabla} \phi_{j})^2(\mathbf{x})+\frac{1}{2} m_0^2 \phi_{j}(\mathbf{x})^2+\frac{\lambda_0}{4!}\phi_{j}(\mathbf{x})^4\right)} \\
        =& \left(\sqrt{\frac{a}{2\pi a_0}}\right)^{\mathcal{V}} \left(\frac{a}{a_0}-\frac{a^2}{a_0^2} \left(\phi_{j+1}(\mathbf{x}_k)-\phi_{j}(\mathbf{x}_k)\right)^2\right) e^{-\frac{a}{a_0} \sum_{\mathbf{x}}\frac{1}{2}\left(\phi_{j+1}(\mathbf{x})-\phi_{j}(\mathbf{x})\right)^2-\frac{a_0}{a}\sum_{\mathbf{x}}  \left(\frac{1}{2}(\mathbf{\nabla} \phi_{j})^2(\mathbf{x})+\frac{1}{2} m_0^2 \phi_{j}(\mathbf{x})^2+\frac{\lambda_0}{4!}\phi_{j}(\mathbf{x})^4\right)},
    \end{aligned}
    \label{eqn:path_int_scalar_pi2_deriv}
\end{equation}
where we have shifted $\pi(\mathbf{x})\rightarrow \pi(\mathbf{x})+i \frac{a}{a_0}\left(\phi_{j+1}(\mathbf{x})-\phi_{j}(\mathbf{x})\right)$ in the second-to-last equality and we have used the identities $\int_{-\infty}^{+\infty} \dd x \; x e^{-x^2/2\sigma^2}=0$ and $\int_{-\infty}^{+\infty} \dd x \; x^2 e^{-x^2/2\sigma^2}=\sigma^3 \sqrt{2\pi}$ to derive the final line.
Hence, in a path integral, to evaluate $\hat \pi(\mathbf{x}_k)^2$ at time slice $t$, one can replace it by
\begin{equation}
    \frac{a}{a_0}-\frac{a^2}{a_0^2} \left(\phi(t+a_0,\mathbf{x}_k)-\phi(t,\mathbf{x}_k)\right)^2.
    \label{eqn:path_int_scalar_pi2_expr}
\end{equation}
From Eqns. \ref{eqn:path_int_scalar_O_phi} and \ref{eqn:path_int_scalar_pi2_expr}, one can derive a path integral expression for the ground state energy $E_0$ of $\hat{H}$ as
\begin{equation}
    E_0 \approx \frac{\Tr[\hat{H}  e^{-\hat H T}]}{    \Tr[e^{-\hat H T}]}
    =  \frac{\int  \mathcal{D}\phi  \tilde{H}[\phi]e^{-S[\phi]}}{\int  \mathcal{D}\phi  e^{-S[\phi]}} ,
\end{equation}
where $S[\phi]$ is given in Eqn.~\ref{eqn:path_int_scalar_S_unmodified} and 
$\tilde{H}[\phi]$ denotes the expression in terms of $\phi$ corresponding to the Hamiltonian operator $\hat H$, defined as
\begin{equation}
 \tilde{H}[\phi]=   \frac{1}{a} \sum_{\mathbf{x}}\left( \frac{1}{2} \left(
\frac{a}{a_0} - \frac{a^2}{a_0^2} (\phi (t + a_0, \mathbf{x})
- \phi (t, \mathbf{x}))^2 \right) + \frac{1}{2}
(\mathbf{\nabla} \phi)^2 (t, \mathbf{x}) +
\frac{1}{2} m_0^2 \phi (t, \mathbf{x})^2 +
\frac{\lambda_0}{4!} \phi (t, \mathbf{x})^4\right),
\label{eqn:path_int_scalar_Htilde_unmodified}
\end{equation}

The ground state of a modified Hamiltonian can be obtained analogously. For example, the vacuum energy  $E'_{0,\phi(\mathbf{x_i})}$ of the Hamiltonian $\hat H'_{\phi(\mathbf{x}_i)}$ can be obtained from
\begin{equation}
    E'_{0,\phi(\mathbf{x_i})} \approx \frac{\int  \mathcal{D}\phi  \tilde{H}'_{\phi(\mathbf{x_i})}[\phi]e^{-S'_{\phi(\mathbf{x_i})}[\phi]}}{\int  \mathcal{D}\phi  e^{-S'_{\phi(\mathbf{x_i})}[\phi]}},
\end{equation}
and $S'_{\phi(\mathbf{x_i})}[\phi]$ and $\tilde{H}'_{\phi(\mathbf{x_i})}[\phi]$ are given as
\begin{equation}
    S'_{\phi(\mathbf{x_i})}[\phi] = S[\phi] - \sum_t\frac{a_0}{a} \frac{1}{2} m_0^2 \phi(t, \mathbf{x}_i)^2,\label{eqn:path_int_scalar_S_phi_x_i}
\end{equation}
and
\begin{equation}
    \tilde{H}'_{\phi(\mathbf{x_i})}[\phi] = \tilde{H}[\phi]-\frac{1}{a}\frac{1}{2} m_0^2  \phi(t, \mathbf{x}_i)^2.
    \label{eqn:path_int_scalar_Htilde_phi_x_i}
\end{equation}

\subsection{Path integral in 2+1D U(1) gauge theory in dual formalism \label{app:path_integral:dual} }
The path integral for the dual formalism of 2+1D U(1) gauge theory can be derived analogously. Here, we will derive a path integral representation of $\Tr[e^{-\hat H_{\text{U(1)}} T}]$ and $\Tr[\hat O e^{-\hat H_{\text{U(1)}} T}]$. Once these quantities are derived, the corresponding path integral for the modified Hamiltonian $\hat H'_{B^{(\infty)},\eta}$  can be derived straight-forwardly, since $\hat H'_{B^{(\infty)},\eta}$ differs from $\hat H_{\text{U(1)}}$ only by a potential energy term expressed in terms of $\hat{B}$ field. The path integral corresponding to $\hat H'_{R_p, \eta}$ can also be easily obtained simply by replacing the matrix $M_{r,s}$ by $M'_{r,s}\equiv M_{r,s}-\eta \delta_{r,p} \delta_{p,s}$. 

To derive $\Tr[e^{-\hat H_{\text{U(1)}} T}]$ and $\Tr[\hat O e^{-\hat H_{\text{U(1)}} T}]$, it is convenient to rewrite Eqn. \ref{eqn:dual_hamiltonian} in another form. Note, one can diagonalize the matrix $M_{r,s}$ as $M_{r,s} =\sum_i U_{r,i} \lambda_i U^\dagger_{i,s}$.  Since $M_{r,s}$ is a positive-definite real symmetric matrix, all eigenvalues $\lambda_i$ are real and positive, and the unitary matrix $U_{r,i}$ can be shown to be real and orthogonal, meaning $U^T=U^\dagger$. Then, one can define $\hat S_i = \sum_r U^T_{i,r} \hat R_r $ and $\hat C_i = \sum_r U^T_{i,r} \hat B_r$. By the orthonormal property of $U$, one can show that $[\hat C_i, \hat S_j] = i \delta_{i,j}$. Using the new variables, the Hamiltonian in Eqn.~\ref{eqn:dual_hamiltonian} can be written as 
\begin{equation}
      \hat H_{\text{U(1)}} = \frac{1}{a} \frac{g^2}{2} \sum_{i} \lambda_i \hat S_i^2+\frac{1}{a}\frac{1}{2 g^2} \sum_{i} \left(\hat C_{i}\right)^2.
\end{equation}
Having written $ \hat H_{\text{U(1)}} $ in a form that is similar to the scalar field theory case, one can use a similar approach and express the action as 
\begin{equation}
    S_{\text{U(1)}} = 
    \sum_{t,i}\left(\frac{a}{a_0} \frac{1}{2g^2}\frac{1}{\lambda_i}\left(C_i(t+a_0)-C_i(t)\right)^2+\frac{a_0}{a}\frac{1}{2g^2} C_i(t)^2\right).
\end{equation}
Then, plugging back $C_i = \sum_r U^T_{i,r} B_r$, the action can be written as
\begin{equation}
     S_{\text{U(1)}} = \sum_t \left(\frac{a}{a_0} \frac{1}{2g^2}\sum_i \frac{\left( \sum_p U^T_{i,p}\left(B_p(t+a_0) -B_p(t)\right)\right)^2}{\lambda_i}+\frac{a_0}{a}\frac{1}{2g^2}\sum_p (B_p)^2 \right).
\end{equation}
The action for the 2+1D non-compact U(1) gauge theory can thus be obtained. With minimal modifications, Eqns.~\ref{eqn:dual_S_Bmax} and \ref{eqn:dual_R_action} can be similarly obtained. 

Also, similar to how $\bra{\Omega} \hat \pi(\mathbf{x}_k)^2 \ket{\Omega}$ in scalar field theory is evaluated by the path integral using Eqn. \ref{eqn:path_int_scalar_pi2_expr}, the value $\bra{\Omega} \hat S_i^2 \ket{\Omega}$ can be evaluated in a path integral, where $\hat S_i^2$ will be replaced by the expression
\begin{equation}
     \frac{a}{a_0 g^2 \lambda_i}- \left(\frac{a\left(C_i(t+a_0)-C_i(t)\right)}{a_0 g^2 \lambda_i}\right)^2.
     \label{eqn:path_int_dual_S2_expr}
\end{equation}
If one wishes to evaluate $\bra{\Omega} \hat R_p^2 \ket{\Omega}$, notice
\begin{equation}
    \begin{aligned}
        \bra{\Omega} \hat R_p^2 \ket{\Omega} =&  \bra{\Omega} ( \sum_i U_{p,i} \hat S_i)^2 \ket{\Omega} \\
        =&  \sum_i (U_{p,i})^2\bra{\Omega}  \hat S_i^2 \ket{\Omega} + \sum_{i\neq j} U_{p,i} U_{p,j} \bra{\Omega} \hat S_i \hat S_j \ket{\Omega}.
        % =&  \sum_i (U_{p,i})^2\bra{\Omega}  \hat S_i^2 \ket{\Omega} ,
    \end{aligned}
\end{equation}
From Eqn.~\ref{eqn:path_int_dual_S2_expr}, one can see that $\sum_i (U_{p,i})^2\bra{\Omega}  \hat S_i^2 \ket{\Omega}$ corresponds to the following quantity in the path integral:
\begin{equation}
    \sum_i (U_{p,i})^2 \left(  \frac{a}{a_0 g^2 \lambda_i}- \left(\frac{a\sum_q U^T_{i,q}\left(B_q(t+a_0)-B_q(t)\right)}{a_0 g^2 \lambda_i}\right)^2\right).
    \label{eqn:path_int_dual_R2_diag}
\end{equation}
For the expectation value $\bra{\Omega} \hat S_i \hat S_j \ket{\Omega}$, we need to derive an expression corresponding to $\hat S_i \hat S_j$. In a similar spirit to the derivation of Eqn.~\ref{eqn:path_int_scalar_pi2_deriv}, we analyze $\bra{\mathbf{B}(t+a_0)}\hat S_i \hat S_j e^{-\hat H a_0}\ket{\mathbf{B}(t)} $:
\begin{equation}
    \begin{aligned}
        & \bra{\mathbf{B}(t+a_0)}\hat S_i \hat S_j e^{-\hat H a_0}\ket{\mathbf{B}(t)} \\
        = & \int \mathcal{D}\mathbf{S}    \bra{\mathbf{B}(t+a_0)}\ket{\mathbf{S}}\bra{\mathbf{S}}\hat S_i \hat S_j e^{-\hat H a_0}\ket{\mathbf{B}(t)} \\
        \approx & \int \mathcal{D}\mathbf{S} \left(\frac{1}{2\pi}\right)^{\mathcal{V}} S_i S_j \exp \left( - \frac{g^2 a_0}{2 a} \sum_k \lambda_k S_k^2 -
\frac{a_0}{2 g^2 a} \sum_p (B_p)^2 + i \sum_k S_k (C_k(t+a_0) -
C_k (t) ) \right) \\
= & \int \mathcal{D}\mathbf{S} \left(\frac{1}{2\pi}\right)^{\mathcal{V}} S_i S_j \exp \left( - \frac{g^2 a_0}{2 a} \sum_k \lambda_k \left(S_k-i\frac{a\Delta C_k}{a_0 g^2 \lambda_k}\right)^2 -\frac{a}{2 a_0 g^2}\sum_k \frac{\Delta C_k^2}{\lambda_k}-
\frac{a_0}{2 g^2 a} \sum_p (B_p)^2 \right) \\
= & \int \mathcal{D}\mathbf{S} \left(\frac{1}{2\pi}\right)^{\mathcal{V}} \left(S_i+i\frac{\Delta C_i}{a_0 g^2 \lambda_i}\right)\left(S_j+i\frac{\Delta C_j}{a_0 g^2 \lambda_j}\right)\exp \left( - \frac{g^2 a_0}{2 a} \sum_k \lambda_k S_k^2 -\frac{a}{2 a_0 g^2}\sum_k \frac{\Delta C_k^2}{\lambda_k}-
\frac{a_0}{2 g^2 a} \sum_p (B_p)^2 \right) \\
=&  \left(\prod_{k=1}^{\mathcal{V}}\sqrt{\frac{a}{2\pi a_0 g^2 \lambda_k}}\right) \left(-\frac{a^2\Delta C_i \Delta C_j}{a_0^2 g^4 \lambda_i \lambda_j}\right) \exp \left(-\frac{a}{2 a_0 g^2}\sum_k \frac{\Delta C_k^2}{\lambda_k}-
\frac{a_0}{2 g^2 a} \sum_p (B_p)^2 \right),
    \end{aligned}
\end{equation}
where $\Delta C_i$ denotes $C_i(t+a_0) -
C_i (t)$, with $C_i =\sum_r U^T_{i,r} B_r $. Thus, to evaluate the value of $\hat S_i \hat S_j$, one needs to evaluate the expression
$$-\frac{a^2\Delta C_i \Delta C_j}{a_0^2 g^4 \lambda_i \lambda_j},$$
in a path integral, which can be expanded as
\begin{equation}
   -\left(\frac{a \sum_q (U)^T_{i,q}(B_q(t+a_0) -B_q(t))}{a_0 g^2 \lambda_i}\right)\left(\frac{a \sum_r (U)^T_{j,r}(B_r(t+a_0) -B_r(t))}{a_0 g^2 \lambda_j}\right).\label{eqn:path_int_dual_R2_off}
\end{equation}
Putting together Eqns.~\ref{eqn:path_int_dual_R2_diag} and \ref{eqn:path_int_dual_R2_off}, one can see that $\bra{\Omega} \hat R_p^2 \ket{\Omega}$ can be evaluated in a path integral by evaluating the value of
\begin{equation}
 \sum_{i,j} U_{p,i} U_{p,j} \left( \frac{a \delta_{i,j}}{a_0 g^2 \lambda_i}-\left(\frac{a \sum_q (U)^T_{i,q}(B_q(t+a_0) -B_q(t))}{a_0 g^2 \lambda_i}\right)\left(\frac{a \sum_r (U)^T_{j,r}(B_r(t+a_0) -B_r(t))}{a_0 g^2 \lambda_j}\right)\right).
    \label{eqn:path_int_dual_R2_expr}
\end{equation}
As mentioned earlier, $\hat H'_{R_p, \eta}$ can be obtained from $\hat H_{\text{U(1)}}$ by replacing $M_{r,s}$ by $M'_{r,s}$. Thus, similar to Eqn.~\ref{eqn:path_int_dual_R2_expr}, Eqn.~\ref{eqn:dual_R_R2tilde} can be obtained.

\section{Extrapolation to zero temporal lattice spacing \label{app:extrapolate}}

\begin{figure}[h!]
    \centering
    \includegraphics[width=0.5\linewidth]{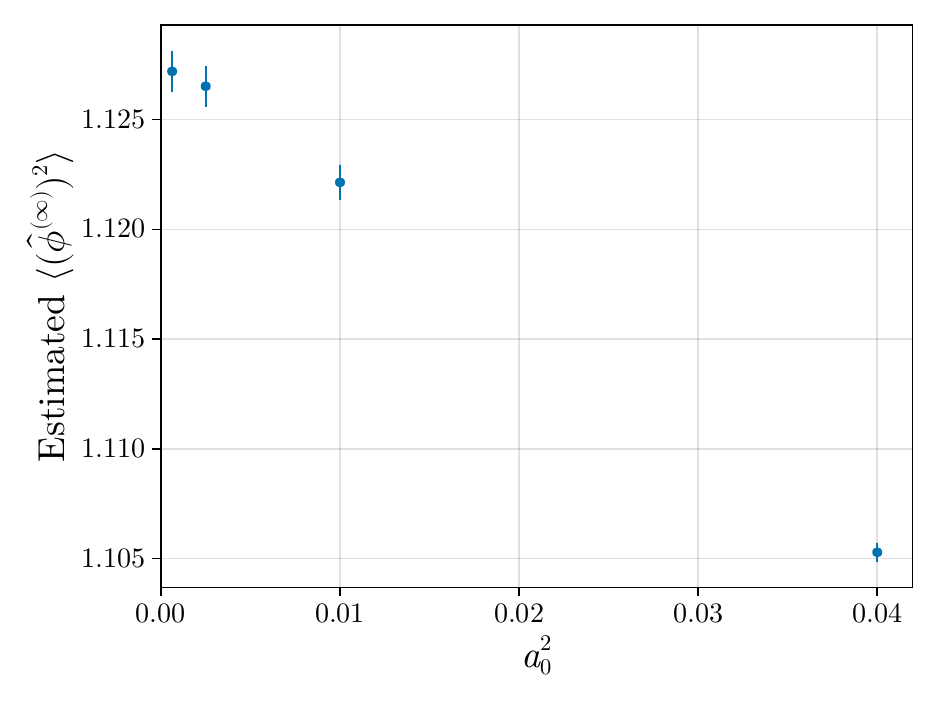}
    \caption{Extrapolation towards zero temporal lattice spacing $a_0$ for the $\phi^4$ scalar field theory. The expectation value $\bra{\Omega_{\phi^{(\infty)},\eta}}(\hat \phi^{(\infty)})^2\ket{\Omega_{\phi^{(\infty)},\eta}}$ for $N_S=16$ is evaluated at different temporal lattice spacings $a_0$. In the plot, $a_0$ is expressed in unit of $a$. The bare parameters are set to be $m_0=1/2$ and $\lambda_0=16$ in lattice unit $a$.
}
    \label{fig:scalar_a0_m0p5_lambda16_ns16}
\end{figure}

As mentioned in Appendix~\ref{app:path_integral}, the path integral reproduces observable values in the limit where the temporal lattice spacing $a_0 \rightarrow 0$. In numerical computation, only finite value of $a_0$ is possible. Therefore, to get precise answer, one needs to evaluate the relevant physical quantities at different values of $a_0$ and extrapolate the result to zero temporal lattice spacing. As an illustration, we showcase in Fig. \ref{fig:scalar_a0_m0p5_lambda16_ns16} the measured results for $\bra{\Omega_{\phi^{(\infty)},\eta}}(\hat \phi^{(\infty)})^2\ket{\Omega_{\phi^{(\infty)},\eta}}$ for $N_S=16$ at various temporal lattice spacings. Here, the bare parameters are set to be $m_0=1/2$ and $\lambda_0=16$ in lattice units. As shown in the figure, at $a_0$ as large as $0.2a$, the result appears to differ considerably from the continuum result; however, at $a_0=0.05a$ or lower, the results appear to be reasonably close to the continuum limit. In principle, one can perform a fit and extracts the $a_0\rightarrow 0$ limit from the fit. 

\begin{figure}[h!]
    \centering
    \includegraphics[width=0.5\linewidth]{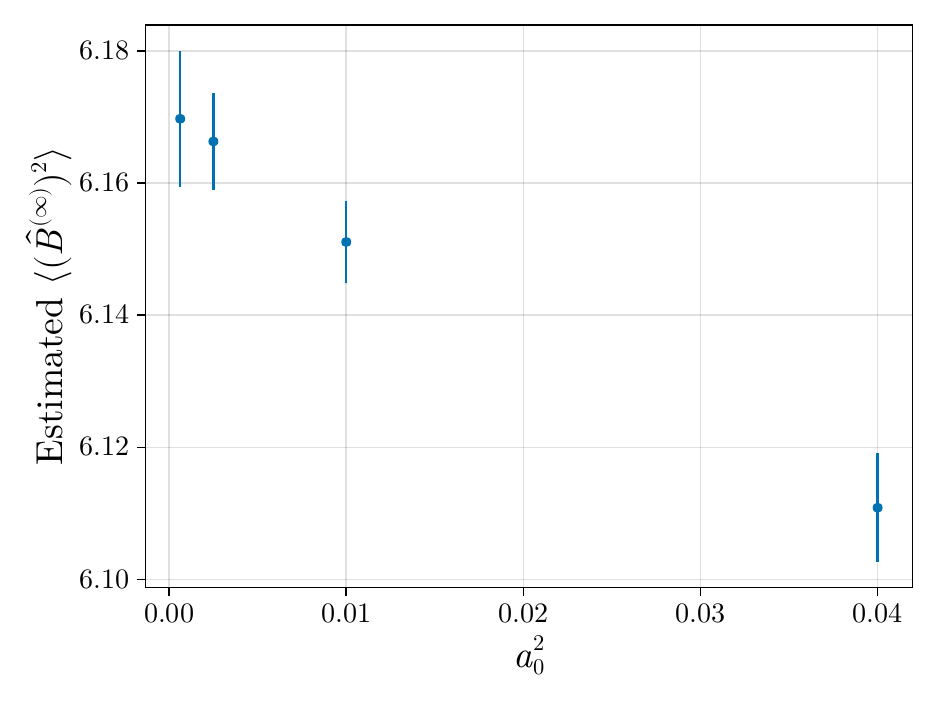}
    \caption{Extrapolation towards zero temporal lattice spacing $a_0$ for the 2+1D U(1) gauge theory in dual formalism. The expectation value for $\bra{\Omega_{B^{(\infty)},\eta}} ( \hat B^{(\infty)} )^2 \ket{\Omega_{B^{(\infty)},\eta}}$ for $N_S=6$ is evaluated at different temporal lattice spacings $a_0$. In the plot, $a_0$ is expressed in unit of $a$.  The bare coupling is set to $g=1$ in lattice unit $a$}
    \label{fig:dual_a0_g1_ns6}
\end{figure}

Likewise, we present the zero $a_0$ extrapolation for the 2+1D non-compact U(1) gauge theory in dual formalism in Fig.~\ref{fig:dual_a0_g1_ns6}. We compute the expectation values $\bra{\Omega_{B^{(\infty)},\eta}} ( \hat B^{(\infty)} )^2 \ket{\Omega_{B^{(\infty)},\eta}}$ at different temporal lattice spacings $a_0$. Here, the bare coupling $g$ is set to $1$ and lattice size $N_S$ set to $6$. As shown in the figure, the value evaluated at larger $a_0$ seems to deviate considerably from the continuum limit, while the expectation values appear to be closer to this limit at small values of $a_0$. One can perform a fit to extrapolate the zero $a_0$ limit.

\FloatBarrier

\bibliography{refs.bib}
\end{document}